\documentclass[%
amsmath,amssymb,
aps,
pra,
twocolumn,
floatfix,
]{revtex4-2}

\usepackage{graphicx}
\usepackage{dcolumn}
\usepackage{bm}
\usepackage{physics}
\usepackage{xcolor}
\usepackage{adjustbox}
\usepackage{xparse}
\usepackage{mathtools}
\usepackage{tikz}
\usetikzlibrary{quantikz2}
\usepackage{amsthm}
\usepackage{comment}
\usepackage{hyperref}

\newcommand{\MMD}{\mathrm{MMD}}
\newcommand{\KL}{\mathrm{KL}}

\newcommand{\muxctrl}[1]{%
    \octrl{#1}%
    \push{\makebox[0pt][c]{\rotatebox{45}{\rule{0.75em}{0.4pt}}}}%
}

\DeclareMathOperator{\diag}{diag}

\makeatletter
\newenvironment{appfigure}
  {\par\addvspace{\intextsep}\noindent
   \begin{minipage}{\linewidth}\def\@captype{figure}\centering}
  {\end{minipage}\par\addvspace{\intextsep}}
\makeatother

\begin{document}

\title{Quantum Fokker-Planck solvers as trainable generative models}

\author{Tariq Mahassen}
\affiliation{School of Mathematical and Physical Sciences, University of Sheffield, Sheffield S3 7RH, United Kingdom}

\author{Marcin Jastrzebski}
\affiliation{School of Mathematical and Physical Sciences, University of Sheffield, Sheffield S3 7RH, United Kingdom}

\author{Oleksandr Kyriienko}
\affiliation{School of Mathematical and Physical Sciences, University of Sheffield, Sheffield S3 7RH, United Kingdom}

\begin{abstract}
We develop and analyze generative models based on quantum solvers of stochastic differential equations. Unlike the common approach based on layered variational quantum circuits with parameterized rotations, we parameterize the potential of a Langevin process. The associated Fokker-Planck (FP) equations are Hermitianized and solved using a linear combination of cosine operators implemented through quantum singular value transformation. By encoding the square root of the stationary density, the model enables sampling through computational-basis measurements and is efficiently controlled by coefficients of the block-encoded diagonal potential. We train the proposed model to match empirical distributions within the family described by FP equations, demonstrating parameter recovery for multivariate probability distributions. Crucially, the model remains trainable at scale. Sample-efficient and convex optimization for in-distribution targets arises from a physics-informed inductive bias and holds regardless of the dimensionality of the problem. We provide circuit constructions and efficient block encodings of the main components of the algorithm and analyze resource requirements. Our approach frames quantum generative learning as quantum model discovery and provides a foundation for extending this framework to high-dimensional, strongly correlated scientific data.
\end{abstract}

\maketitle
\section{Introduction}\label{sec:intro}

Generative modeling relies on learning a probability distribution from data and designing machine learning (ML) models that draw new samples from it~\cite{Goodfellow2016}. Diffusion and flow-based models are now ubiquitous for tasks ranging from image generation to scientific discovery in materials science and high-energy physics~\cite{ho2020denoising, song2021scorebased, rezende2015flows, albergo2019flow}. Quantum machine learning (QML) aims to build models for the same tasks \cite{biamonte2017qml,benedetti2019parameterized,Schuld2022PRXQ}. QML typically relies on parameterized quantum circuits (PQCs)~\cite{mitarai2018learning, cerezo2021variational}, encoding data into quantum states~\cite{schuld2019feature} and treating circuit angles as trainable weights. Generative modeling follows naturally from quantum measurement, which is inherently probabilistic and obeys the Born rule, $x \sim p_{\theta}(x) = |\langle x|\psi_{\theta} \rangle|^2$. Quantum circuit Born machines (QCBMs) use this directly~\cite{liu2018differentiable, coyle2020born}. An extra motivation comes from the complexity of random circuit sampling~\cite{hangleiter2023sampling} and from a predicted quantum advantage for specific generative tasks~\cite{huang2025advantage}. Other model families include hybrid adversarial networks~\cite{zoufal2019generative, sedrakyan2024photonic}, quantum Boltzmann machines and mixed-state models~\cite{amin2018boltzmann, zoufal2021boltzmann, demidik2025equivalence, wilde2025generative, wilde2025hidden}. Spectral models have gained attention for being inherently regularized in the latent space of basis functions~\cite{belis2026spectral}, with sampling enabled by a fixed transform to the computational basis~\cite{kyriienko2024protocols, williams2023chebyshev, wu2025hartley, tuysuz2026fourier}. Multivariate and correlated distributions have been addressed with joint-distribution circuits~\cite{zhu2022joint, barthe2025multivariate,rehm2026hep, martinez2025fragmentation} and physics-based models motivated by photonic and hybrid platforms~\cite{gottlieb2026photonic, kurkin2026universality, kolarovszki2026bosonic, bako2025fermionic, wang2026polariton}. 
\begin{figure*}[t!]
    \centering
    \includegraphics[width=\linewidth]{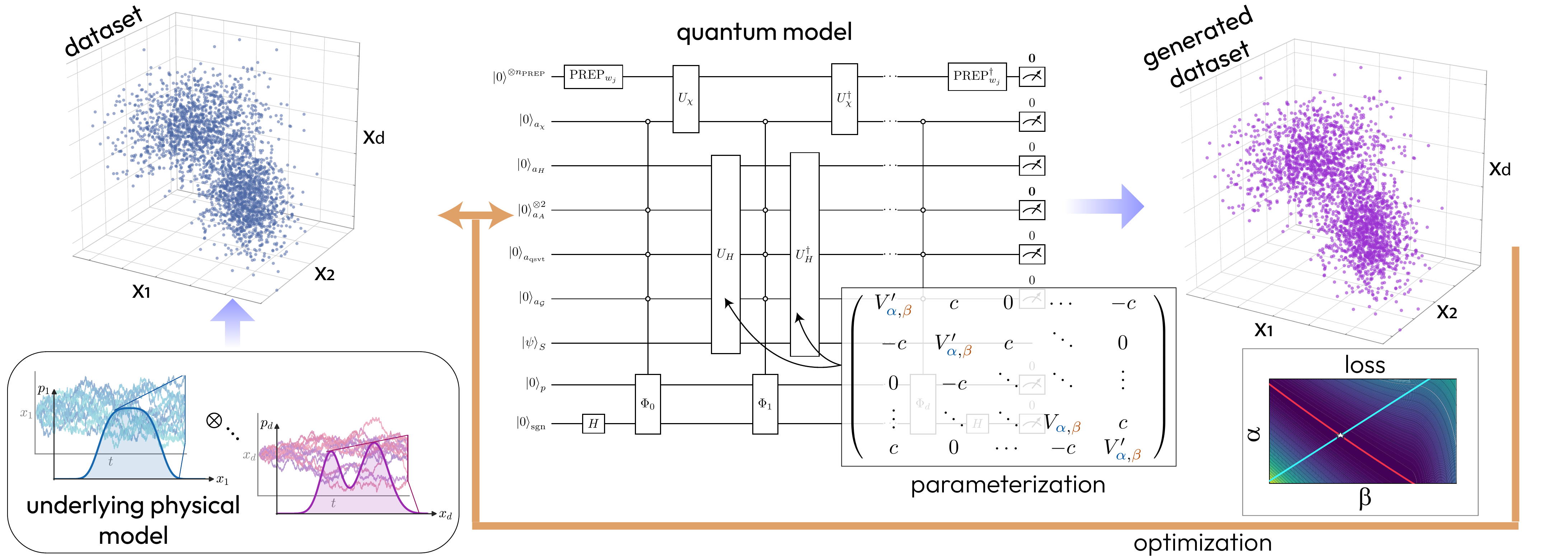}
    \caption{Overview of the proposed approach for physics-informed parameterization of quantum generative models and matching empirical distributions. A target physical model leads to stationary-state samples, which constitute a dataset. The trainable architecture uses a highly structured circuit with a small number of trainable parameters (indicated by $\alpha$, $\beta$), which are naturally reuploaded multiple times throughout the QSVT sequence. The loss is evaluated to minimize the distance between generated and empirical distributions, effectively representing quantum model discovery.
    }
    \label{fig:overview}
\end{figure*}

The performance of QML models largely depends on architectural choices and their parameterization. The classical recipe of adding depth and parameters \cite{Belkin2019} does not transfer directly to quantum circuits \cite{larocca2023theory}. Sufficiently expressive, unstructured circuits exhibit barren plateaus, where gradients vanish exponentially in the number of qubits and training from a polynomial number of measurements fails~\cite{mcclean2018barren, holmes2022connecting, larocca2025barren}. The severity depends on the architecture, on the loss and its estimation~\cite{cerezo2021cost, rudolph2024trainability}, and on the initialization~\cite{lerch2026iqp}, while training shallow circuits can suffer from multiple local minima~\cite{anschuetz2022beyond}. When training on quantum hardware, gradient estimation from finite samples becomes expensive as the number of parameters grows~\cite{Schuld2019parametershift, kyriienko2021differentiation, demidik2025sample}. These barriers have motivated shallow and problem-informed circuits~\cite{benedetti2019shallow, bako2026graphical}, modified training approaches~\cite{demidik2025sample}, and protocols that train classically and reserve the quantum device for sampling~\cite{kasture2023classically, recio2025train, huang2026spectral, tuysuz2026fourier, herrero2025ultimatum}. The tension between trainability guarantees and classical simulability~\cite{cerezo2025does} further suggests changing the model parameterization and untying the number of parameters from the number of qubits~\cite{kyriienko2025discrete}. 

Quantum algorithms for differential equations offer another route to preparing states and probability distributions. This can be done at a cost polylogarithmic in the grid size, under assumptions on operator access, conditioning, and state preparation~\cite{morales2026survey, berry2014high, childs2021high, williams2026iterative}. Non-unitary dynamics can be treated by linear combinations of Hamiltonian simulation or by Schr\"odingerization~\cite{an2023lchs, An_2025lchs, jin2024schrodingerization, jin2023technical}. Nonlinear problems have been approached through Carleman and Kolmogorov linearization \cite{liu2021nonlinear,bravyi2026stochastic}, latent-space physics-informed learning \cite{kyriienko2021solving, kyriienko_piqml}, and effective Hamiltonians whose ground state encodes the solution~\cite{Subasi2019,Jennings2025,wu2025effective}. Stochastic differential equations are a natural target, since the density of their trajectories obeys a Fokker--Planck equation~\cite{oksendal1998sde}. Quantum algorithms for stochastic dynamics and Fokker--Planck equations have been developed~\cite{kubo2021variational, alghassi2022variational, an2021quantumaccelerated, garciamolina2022fourier, jin2024fokker, tennie2025fp, kharazi2026provablequantumspeedupsreactionratelchs, kyriienko2024protocols, paine2021quantile, wang2025diffusion}. While reading out a full amplitude-encoded solution can consume the speedup~\cite{aaronson2015read}, sampling is a natural target that avoids full reconstruction. At the same time, a solver for a given operator is not yet a learned generative model.

In this work we construct such a model. The trainable parameters are the coefficients of the potential of an overdamped Langevin process, and the circuit is a quantum solver that prepares the stationary distribution of the associated Fokker--Planck equation. Hermitianization of the Fokker--Planck generator gives a positive semidefinite operator whose ground state is the square root of the stationary density. We prepare this state with an imaginary-time filter implemented as a linear combination of cosine operators through quantum singular value transformation (QSVT)~\cite{gilyen2019quantum, martyn2021grand, an2023lchs}. The potential enters through a parameterized block encoding, so the number of trainable parameters is fixed by the chosen dictionary of potential terms and is independent of the grid resolution, while the prepared state spans an exponentially large configuration space. We show that the model recovers the parameters of multivariate, correlated distributions within its family. The stationary density belongs to an exponential family, and its training reduces to matching a small number of expectation values. The learned coefficients define a stochastic model consistent with the data, which places the approach alongside classical and quantum model discovery~\cite{brunton2016discovering, boninsegna2018sparse, karniadakis2021physics, heim2021discovery}.


\section{Fokker--Planck generative model}\label{sec:model}

We start by considering a $d$-dimensional overdamped Langevin process with additive, isotropic noise, $ d\mathbf{X}_t = -\nabla V(\mathbf{X}_t)\,dt + \sigma\, d\mathbf{W}_t$, where $\mathbf{W}_t$ is a standard $d$-dimensional Wiener process, $\sigma>0$ is a noise strength, and $V:\mathbb{R}^d\to\mathbb{R}$ is a confining potential. The deterministic drift $-\nabla V$ pushes trajectories toward local minima of $V$. The noise adds fluctuations and allows particles to cross barriers. The probability density $p(\mathbf{x},t)$ of an ensemble of such trajectories obeys the Fokker--Planck equation~\cite{risken1989}
\begin{equation}
\label{eq:fp}
    \partial_t p = \nabla\cdot\big(p\,\nabla V\big) + \tfrac{\sigma^2}{2}\,\Delta p
    = -\nabla\cdot\mathbf{J},
\end{equation}
where $\mathbf{J} = -p\nabla V -\tfrac{\sigma^2}{2}\nabla p$ is the probability current.

Imposing the equilibrium condition $\mathbf{J}=0$ gives $\nabla\log p_{\rm ss} = -2\nabla V/\sigma^2$ and hence the stationary density $ p_{\rm ss}(\mathbf{x}) = Z^{-1} \exp\!\left(-2V(\mathbf{x})/\sigma^2\right)$, where $Z = \int e^{-2V(\mathbf{x})/\sigma^2}\, d\mathbf{x}$.
The stationary distribution concentrates near the minima of $V$, with $\sigma$ controlling both its width and the rate at which probability crosses barriers between minima.

We now let the potential carry the trainable parameters. We take $ V(\mathbf{x};\boldsymbol\theta) = \sum_{j=1}^{K} \theta_j\, v_j(\mathbf{x})$
with a fixed dictionary of basis functions $\{v_j\}$, in this work monomials in the coordinates, and coefficients $\boldsymbol\theta\in\mathbb{R}^K$. Substituting into the stationary density $p_{\mathrm{ss}}$, the model distribution is
\begin{equation}\label{eq:model_density}
    p_{\boldsymbol\theta}(\mathbf{x})
    = \frac{1}{Z(\boldsymbol\theta)}
      \exp\!\Big(-\frac{2}{\sigma^2}V(\mathbf{x};\boldsymbol{\theta})\Big).
\end{equation}
This is a generative model in the usual sense: given $\boldsymbol\theta$ it defines a distribution, and given samples from an unknown distribution we can fit $\boldsymbol\theta$ so that $p_{\boldsymbol\theta}$ matches them. It is also a physical model: the fitted $\boldsymbol\theta$ are the coefficients of the drift of a stochastic differential equation consistent with the data, which is why we link the training task to model discovery. If the data are indeed generated by an overdamped Langevin process with a potential in the span of the dictionary, the potential is identifiable from $p_{\rm ss}$ up to an additive constant, $V = -\tfrac{\sigma^2}{2}\log p_{\rm ss} + C$. We assume throughout that $\sigma$ is known or estimated separately from trajectory data. Equivalently, if unknown, we can absorb $\sigma$ into the definition of the trainable parameters $\boldsymbol{\theta}$ and interpret it as aiming to fit the natural parameters $\eta_j^* = -2\theta^*_j/\sigma^2$. These cannot, on their own, identify the process in full, but can form a crucial part of an extended model-discovery pipeline.

Two features of Eq.~\eqref{eq:model_density} are crucial for this model. First, it belongs to an exponential family with natural parameters $-2\theta_j/\sigma^2$ and sufficient statistics $v_j(\mathbf{x})$~\cite{brown1986fundamentals}. This fixes the geometry of likelihood-based training (Sec.~\ref{sec:trainability}). Second, a small $K$ does not mean a simple distribution. A quartic $V(x)=\tfrac{\theta_1}{2}x^2+\tfrac{\theta_2}{4}x^4$ with $\theta_1<0$ is bimodal. In several dimensions, coupling terms such as $x_1x_2$ or $x_1^2x_2^2$ correlate the coordinates, and the stationary density is in general neither a product nor a Gaussian. Adding more dimensions and coupling them with nonlinear terms can lead to complicated probability distributions.


Once discretized, the model distribution becomes a probability mass vector over a grid. We consider the trainable potential $V(\mathbf{x};\boldsymbol\theta)$ of Eq.~\eqref{eq:model_density} with $K$ coefficients. Each of the $d$ coordinates is discretized on $N=2^n$ grid points. The generative model therefore has exactly $K$ trainable parameters, and its output $p_{\boldsymbol\theta}$ is a probability vector with $N^d$ entries. The quantum register encoding $\sqrt{p_{\boldsymbol\theta}}$ as amplitudes contains $dn=d\log_2 N$ qubits. The parameter count $K$ is independent of $N$. For a dictionary of monomials of total degree at most $r$, it satisfies $K\le\binom{d+r}{r}-1$.

Our approach separates two resources: the number of trainable parameters and the size of the represented probability vector. In a conventional quantum circuit Born machine with comparable expressivity, these resources are coupled through the ansatz depth. Here, the trainable-parameter count is determined by the dictionary defining the potential, while the probability-vector size is determined by the discretization. The solver connects this compact parameterization to quantum state preparation, with a computational cost discussed in Sec.~\ref{sec:resources}. Throughout this paper, ``efficient parameterization'' refers to the trainable-parameter count. At the same time, the gate count and circuit complexity of our approach are higher than for a QCBM, and hence it is more suitable for fault-tolerant devices.
\begin{figure}[b!]
    \centering
    \includegraphics[width=0.9\columnwidth]{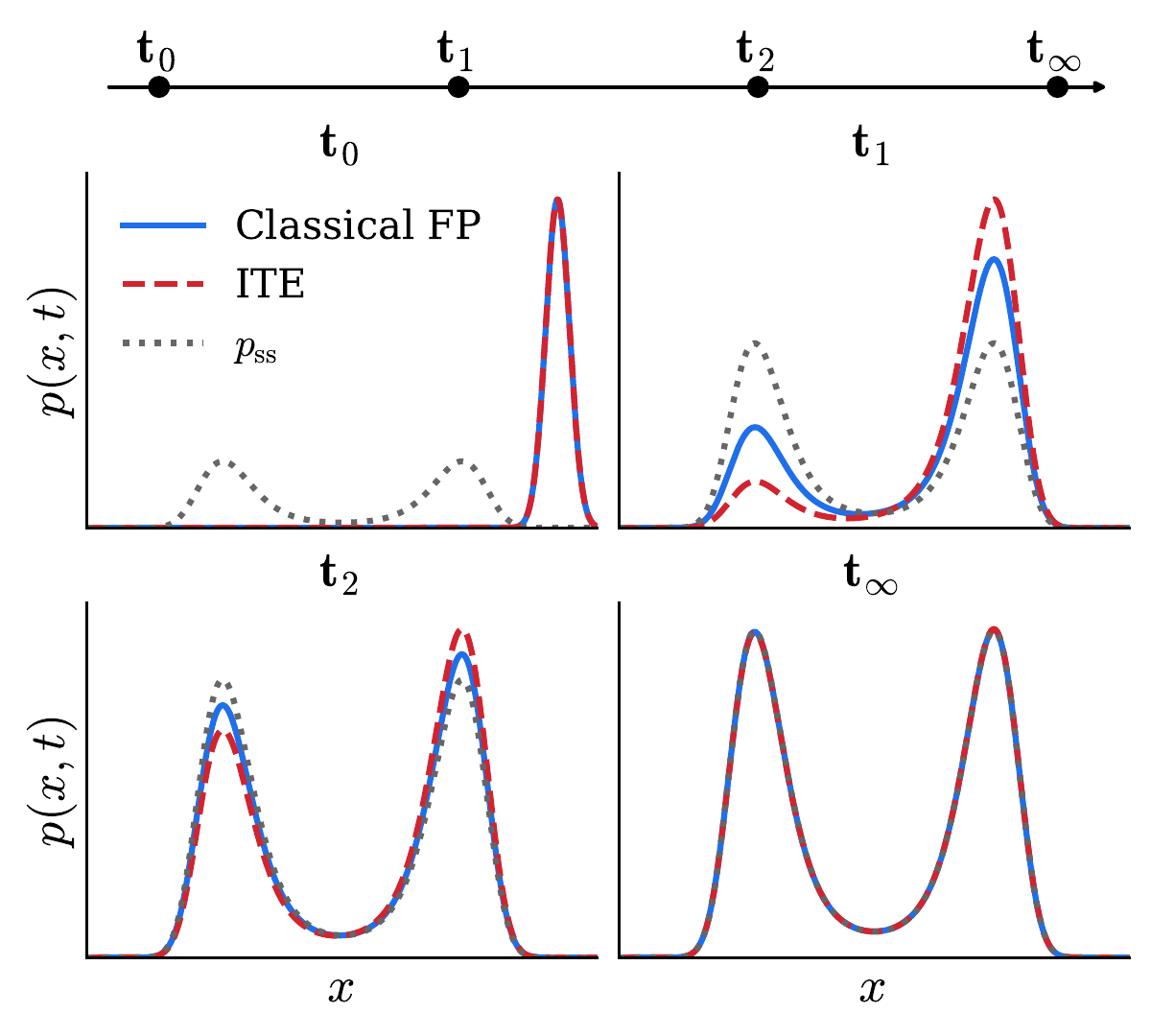}
    \caption{Classical Fokker--Planck evolution (solid blue curves) vs imaginary-time evolution (ITE) (dashed red curves) for the distribution generated by the double-well potential $V(x)=\lambda\left(x^2-a^2\right)^2$ (dotted gray curve). Both start from the same narrow Gaussian at $t_0$. At the intermediate times $t_1$ and $t_2$ the two distributions differ, since the dynamics are not equivalent away from stationarity. At late times ($t_{\infty}$) both relax to the same stationary distribution $p_{\mathrm{ss}}$.}
    \label{fig:ite_vs_fp}
\end{figure}
%


Finally, sampling from Eq.~\eqref{eq:model_density} classically requires either evaluating $Z(\boldsymbol\theta)$ on the $N^d$ grid, which is exponential in $d$, or running a Markov chain, whose mixing time grows when $\sigma$ is small and $V$ has several well-separated minima \cite{giles2015multilevel, beck2016rare}. A quantum register of $d\log_2 N$ qubits holds the discretized $\sqrt{p_{\boldsymbol\theta}}$ as amplitudes, and computational-basis measurement produces a sample from the joint distribution in one shot, with marginals obtained by discarding registers. The task of the quantum algorithm is therefore to prepare the stationary state of the discretized Fokker--Planck operator for a given $\boldsymbol\theta$. Because Eq.~\eqref{eq:fp} is a non-Hermitian relaxation, we solve the problem by Hermitianizing the operator and applying an imaginary-time filter built from cosine operators (see Sec.~\ref{sec:quantum}). 

We emphasize that the model generates stationary-state solutions. Imaginary-time evolution (ITE) exponentially suppresses excited-state contributions, projecting onto the ground state of a Hamiltonian that encodes the stationary Fokker--Planck distribution. In contrast, Fokker--Planck dynamics describe the physical time evolution of a probability distribution associated with an ensemble of stochastic trajectories (see Fig.~\ref{fig:ite_vs_fp}), so transient ITE dynamics are not considered.

\section{Quantum implementation}\label{sec:quantum}

The quantum model takes $\boldsymbol\theta$ as input and prepares a state with the discretized Born-rule distribution $p_{\boldsymbol\theta}$ of Eq.~\eqref{eq:model_density} embedded into amplitudes. To build such a model, we split the workflow into four stages: (A) transformation of the Fokker--Planck generator into a positive semidefinite operator $H=A^\dagger A$ whose ground state is $\sqrt{p_{\boldsymbol\theta}}$; (B) a block encoding of the discretized $A$ in which $\boldsymbol\theta$ enters only through the diagonal potential term; (C) preparation of the ground state by an imaginary-time filter $e^{-tH}$ implemented as a linear combination of cosine operators (LCCO); and (D) measurement and a loss that drives the update of $\boldsymbol\theta$.

\subsection{Hermitianization}

The Fokker--Planck generator in Eq.~\eqref{eq:fp} is generally non-Hermitian, so its propagator is non-unitary and cannot be applied directly on a quantum device. The standard remedy is the similarity transformation by the square root of the
stationary density \cite{risken1989, bernstein1984susy} $p(\mathbf{x},t) = e^{-V(\mathbf{x})/\sigma^{2}}\,\psi(\mathbf{x},t)$, which Hermitianizes the equation by casting Eq.~\eqref{eq:fp} into imaginary-time Schr\"odinger form, $\partial_t \psi=-H \psi$, with $H = -\frac{\sigma^2}{2} \nabla^2 +V_{\rm eff}(\mathbf{x})$ and $ V_{\rm eff}(\mathbf{x}) = |\nabla V|^2 /(2 \sigma^2) - \Delta V /2$ (see App.~\ref{app:hermitianization} for the derivation).
Note that $H$ is self-adjoint, and hence solving for $\psi(\mathbf{x},t)$ requires ITE with the operator $\exp(-Ht)$, which remains non-unitary.


\subsection{Imaginary-time evolution via a linear combination of cosines}

We aim to find the stationary state solution of the Hermitianized Fokker--Planck equation $\partial_t \ket{\psi_{\mathrm{ss}}}=-H \ket{\psi_{\mathrm{ss}}} = 0$. Expressing the Hamiltonian in its spectral decomposition $H= \sum_i E_i \ketbra{E_i}{E_i}$, we see that the imaginary-time propagator is $\exp(-Ht) = \sum_i e^{-E_i t} \ketbra{E_i}{E_i}$. For an initial state with non-zero overlap with the ground state we have $\ket{\psi(0)} = \sum_i c_i \ket{E_i}$ such that ITE prepares an (unnormalized) state of the form $\ket{\psi(t)} = \sum_i c_i e^{-E_i t} \ket{E_i}$.  Since $H \succeq 0$, higher energy eigenstates are exponentially suppressed relative to the ground state at rates determined by the spectral gap $\Delta = E_1 - E_0$  and after $t=O(\log(1/\epsilon)/\Delta)$ time the final state is an $\epsilon$-accurate approximation to the ground state $\ket{E_0}$, up to the dependence on the initial ground state overlap. Therefore, retrieving the stationary state solution to the Fokker--Planck equation is equivalent to ground state preparation of the imaginary-time Schr\"odinger equation. To guarantee $H \succeq 0$ we factorize it as $H=A^\dag A$, where, in one dimension,
\begin{equation} \label{eq:A}
    A = \frac{\sigma}{\sqrt{2}}\partial_x + \frac{1}{\sqrt{2}\sigma} V'(x).
\end{equation} 

The non-unitarity of the imaginary-time propagator necessitates the use of the Hubbard--Stratonovich transformation to express the operator as a linear combination of unitaries (LCU)~\cite{Chakraborty2024}. For a Hermitian matrix $B$ it reads
\begin{equation}\label{eq:HS}
    e^{-B^2 t} =
    \frac{1}{\sqrt{\pi}} \int_{-\infty}^{\infty} e^{-\chi^2} e^{-i2\sqrt{t}\,\chi B}\,d\chi.
\end{equation}
We can construct $D$ quadrature nodes $\chi$ and weights $w$ using the Stieltjes procedure and the Golub--Welsch algorithm~\cite{golub1969calculation} to discretize Eq.~\eqref{eq:HS} as
\begin{equation} \label{LCCS}
    e^{-B^2 t} \approx
    \frac{1}{\sqrt{\pi}} \sum_{j=1}^{D} w_j e^{-i2\sqrt{t}\,\chi_jB},
\end{equation}
which can, in general, be implemented with a linear combination of Hamiltonian simulations.

We use the even and odd symmetries of the quadrature weights and nodes $w_j=w_{-j}$ and $ \chi_j=-\chi_{-j}$ to remove imaginary terms from the summation, resulting in a linear combination of cosines of the Hermitian matrix $B$,
\begin{equation}\label{eq:LCCS}
\begin{aligned}
e^{-B^2 t} \approx \frac{2}{\sqrt{\pi}} \sum_{j: \chi_j>0} w_j \cos \left(2 \sqrt{t} \chi_j B\right).
\end{aligned}
\end{equation}
The cosines, being real, even functions, are precisely what standard QSVT is able to implement without costly extensions of the protocol, and the symmetry essentially halves the number of LCU branches needed. In our construction $B$ is the Hermitian dilation of $A$, $B=\ketbra{0}{1}\otimes A^\dagger+\ketbra{1}{0}\otimes A$, so that $B^2=\ketbra{0}{0}\otimes A^\dagger A+\ketbra{1}{1}\otimes AA^\dagger$ and $e^{-tB^2}$ acts as $e^{-tH}$ on the $\ket{0}$ block. 
We discuss how the approximation of Eq.~\eqref{eq:LCCS} is applied to $H$ within the circuit construction in Sec.~\ref{sec:quantum_implementation}.

Gauss--Hermite nodes and weights are symmetric for any $D$, with $\chi_{-j}=-\chi_j$ and $w_{-j}=w_j$, so the imaginary parts cancel exactly. For odd $D$, the rule also contains the node $\chi=0$, which contributes $w_0\cos(0)=w_0 I$. This identity contribution is included automatically in the cosine implementation, provided that the approximating polynomial equals one at zero signal. We take $D$ even for convenience, such that every branch of Eq.~\eqref{eq:LCCS} carries weight $2w_j$.

\subsection{Discretization and block encoding for the quantum circuit}\label{sec:quantum_implementation}
\begin{figure}[t!]
    \centering
    \begin{adjustbox}{max width=0.80\columnwidth,center}
    $\displaystyle
    A
    =
    \frac{\sigma}{2\sqrt{2}\,\Delta x}
    \left(S_{+}-S_{-}\right)
    +
    \frac{1}{\sqrt{2}\sigma}V'(x)
    \longrightarrow
    U_A = \begin{pmatrix}
        A & * \\
        * & *
    \end{pmatrix}
    $
    \end{adjustbox}
    
    \vspace{1em}
    $\left.
    \vcenter{\hbox{
    \begin{adjustbox}{max width=0.90\columnwidth,left}
    \begin{quantikz}[row sep=0.45cm, column sep=0.4cm]
        \lstick{$\ket{0}_{a_{A,1}}$}
        & \gate[wires=2]{\mathrm{PREP}_A}
        & \qw
        & \octrl{3}
        & \octrl{3}
        & \qw
        & \ctrl{2}
        & \gate[wires=2]{\mathrm{PREP}_A^\dagger}
        & \qw
        \\
        \lstick{$\ket{0}_{a_{A,2}}$}
        & \ghost{\mathrm{PREP}_A}
        & \qw
        & \octrl{2}
        & \ctrl{2}
        & \qw
        & \octrl{1}
        & \ghost{\mathrm{PREP}_A^\dagger}
        & \qw
        \\
        \lstick{$\ket{0}_{a_\mathrm{qsvt}}$}
        & \qw
        & \qw
        & \qw
        & \qw
        & \qw
        & \gate[wires=2]{U_{V'}}
        & \qw
        & \qw
        \\
        \lstick{$\ket{\psi}_S$}
        & \qw
        & \gate{\mathrm{QFT}}
        & \gate{U_{+}}
        & \gate{(e^{i\pi})U_{-}}
        & \gate{\mathrm{QFT}^\dagger}
        & \ghost{V(x)}
        & \qw
        & \qw
    \end{quantikz}
    \end{adjustbox}
    }}
    \hspace{-1.0cm}
    \right\}
    \, \displaystyle U_A$
    \caption{LCU block encoding $U_A$ of the discretized operator $A$. The ancilla register $a_A = (a_{A,1}, a_{A,2})$ and $\ket{0}_{a_2}$ prepare the LCU amplitudes through $\mathrm{PREP}_A$. Shift operators $U_{+}$ and $(e^{i\pi})U_{-}$ are implemented in the Fourier basis, giving $S_{+}-S_{-}$. Ancillary register $a_{\textrm{qsvt}}$ is used to perform QSVT to construct $U_{V'}$, a block encoding of the non-unitary potential $V'(x)$. The final $\mathrm{PREP}_A^\dagger$ coherently uncomputes the LCU register, yielding $A/\alpha_A $ in the all-zero ancilla block of $U_A$.}
    \label{fig:block-encoding-A}
\end{figure}
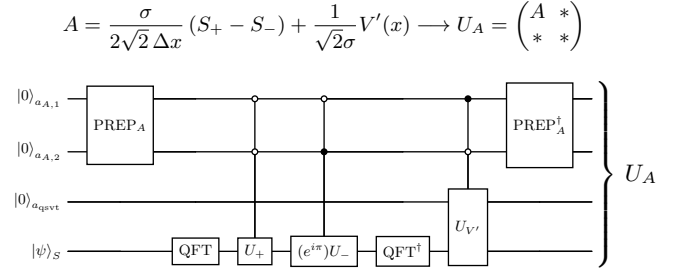

\subsubsection{Block encoding $A$}

To encode $A$ into a quantum circuit we use the central-difference discretization scheme to approximate derivatives as quantum shift operators $S_{\pm}$ (see Ref.~\cite{schalkers2024efficient} for further details). The discretized version of $A$ then becomes a periodic tridiagonal operator, 
\begin{equation}\label{eq:discretized_schrodingerised_fp}
\setlength{\arraycolsep}{2pt}
\renewcommand{\arraystretch}{2.2}
\begin{gathered}
A = \frac{\sigma}{\sqrt{2}} \frac{S_{+}-S_{-}}{2\Delta x}
+ \frac{V'(x_i)}{\sqrt{2}\sigma} \\[1em]
= \frac{1}{\sqrt{2}\sigma}
\left(
\begin{array}{ccccc}
V'(x_0) & c & 0 & \cdots & -c \\
-c & V'(x_1) & c & \ddots & 0 \\
0 & -c & \ddots & \ddots & \vdots \\
\vdots & \ddots & \ddots & V'(x_{N-2}) & c \\
c & 0 & \cdots & -c & V'(x_{N-1})
\end{array}
\right)
\end{gathered}
\end{equation}
where $c = \sigma^2/(2\Delta x)$. To construct this non-unitary operator we first discretize the domain $[-L,L]$ into $N$ grid points $\mathcal{G} = (x_0, \dots, x_{N-1})$ such that $x_i = x_0 + i\Delta x$ where $\Delta x = 2L/N$. We then promote the discretized coordinate vector to the diagonal operator such that $\mathcal{G} = \diag(x_0, \dots, x_{N-1})$. The block encoding $U_\mathcal{G}$ is synthesized using an LCU sequence (see Figs.~\ref{fig:PREP_Xgrid} and \ref{fig:SEL_Xgrid})~\cite{Kharazi2025explicitblock}. We apply a QSVT sequence onto $U_\mathcal{G}$ to implement a polynomial transformation of the diagonal entries, thereby mapping $x_i$ to the desired potential derivative $V'(x_i)$, obtaining $U_{V'}$. The coefficients $\boldsymbol{\theta}$ of the parameterized potential $V(x ; \boldsymbol{\theta})$ are trainable, with the QSVT phase angles $\left\{\phi_k\right\}$ updated accordingly to realize the corresponding polynomial transformation.
The circuits for block encoding $U_\mathcal{G}$ as well as the QSVT sequence to obtain $U_{V'}$ can be found in App.~\ref{app:figures}.

Next, we implement the quantum shift operators $S_{+}$ and $S_{-}$ in Fourier space by conjugating $U_{+}$ and $U_{-}$, respectively, with the quantum Fourier transform (QFT). The negation of the second term requires multiplying $U_{-}$ by the phase factor $e^{i \pi}$. Details of this construction, together with circuit implementations of $U_{ \pm}$, are given in Ref.~\cite{schalkers2024efficient}. We then construct an LCU of the encoded operators $S_{ \pm}$ and $U_{V^{\prime}}$ to obtain the block encoding $U_A$, shown in Fig.~\ref{fig:block-encoding-A}.

\subsubsection{Constructing the signal operator $U_S = U_\chi \otimes U_H$}

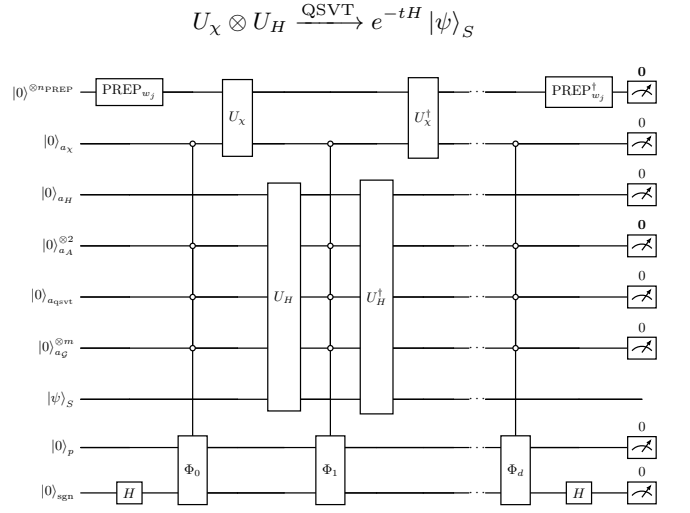
\begin{figure}
    \centering

    \[
    U_\chi \otimes U_H \xlongrightarrow{\mathrm{QSVT}}
    e^{-tH}\ket{\psi}_S
    \]
    
    \begin{adjustbox}{max width=\columnwidth,center}
        \begin{quantikz}[row sep=0.5cm, column sep=0.35cm]
        \lstick{$\ket{0}^{\otimes n_{\rm PREP}}$}
        & \gate{\mathrm{PREP}_{w_j}}
        & \qw
        & \gate[wires=2]{U_\chi}
        & \qw
        & \qw
        & \qw
        & \gate[wires=2]{U_\chi^{\dagger}}
        & \qw
        & \cdots
        & \qw
        & \gate{\mathrm{PREP}_{w_j}^{\dagger}}
        & \meter{\mathbf 0}
        \\
        \lstick{$\ket{0}_{a_\chi}$}
        & \qw
        & \ctrl[open]{4}
        & \ghost{U_\chi}
        & \qw
        & \ctrl[open]{4}
        & \qw
        & \ghost{U_\chi^{\dagger}}
        & \qw
        & \cdots
        & \ctrl[open]{4}
        & \qw
        & \meter{0}
        \\
        \lstick{$\ket{0}_{a_H}$}
        & \qw
        & \qw
        & \qw
        & \gate[wires=5]{U_H}
        & \qw
        & \gate[wires=5]{U_H^{\dagger}}
        & \qw
        & \qw
        & \cdots
        & \qw
        & \qw
        & \meter{0}
        \\
        \lstick{$\ket{0}_{a_A}^{\otimes 2}$}
        & \qw
        & \ctrl[open]{1}
        & \qw
        & \ghost{U_H}
        & \ctrl[open]{1}
        & \ghost{U_H^{\dagger}}
        & \qw
        & \qw
        & \cdots
        & \ctrl[open]{1}
        & \qw
        & \meter{\bf 0}
        \\
        \lstick{$\ket{0}_{a_{\rm qsvt}}$}
        & \qw
        & \ctrl[open]{1}
        & \qw
        & \ghost{U_H}
        & \ctrl[open]{1}
        & \ghost{U_H^{\dagger}}
        & \qw
        & \qw
        & \cdots
        & \ctrl[open]{1}
        & \qw
        & \meter{0}
        \\
        \lstick{$\ket{0}_{a_\mathcal{G}}^{\otimes m}$}
        & \qw
        & \ctrl[open]{2}
        & \qw
        & \ghost{U_H}
        & \ctrl[open]{2}
        & \ghost{U_H^{\dagger}}
        & \qw
        & \qw
        & \cdots
        & \ctrl[open]{2}
        & \qw
        & \meter{0}
        \\
        \lstick{$\ket{\psi}_S$}
        & \qw
        & \qw
        & \qw
        & \ghost{U_H}
        & \qw
        & \ghost{U_H^{\dagger}}
        & \qw
        & \qw
        & \cdots
        & \qw
        & \qw
        & \qw
        \\
        \lstick{$\ket{0}_{p}$}
        & \qw
        & \gate[wires=2]{\Phi_0}
        & \qw
        & \qw
        & \gate[wires=2]{\Phi_1}
        & \qw
        & \qw
        & \qw
        & \cdots
        & \gate[wires=2]{\Phi_d}
        & \qw
        & \meter{0}
        \\
        \lstick{$\ket{0}_{\rm sgn}$}
        & \gate{H}
        & \ghost{\Phi_0}
        & \qw
        & \qw
        & \ghost{\Phi_1}
        & \qw
        & \qw
        & \qw
        & \cdots
        & \ghost{\Phi_d}
        & \gate{H}
        & \meter{0}
        \end{quantikz}
    \end{adjustbox}

    \caption{
    QSVT implementation of the weighted cosine transformation. The block
    encodings $U_\chi$ and $U_H$ are alternated with their adjoints between
    projector-controlled phase operations $\Phi_k$. Each $\Phi_k$ acts on the
    phase ancilla $p$ and sign ancilla $\mathrm{sgn}$, with projector
    $\Pi=\ketbra{0}{0}_{a_\chi}\otimes
    \ketbra{0}{0}_{a_\mathrm{qsvt}} \otimes
    \ketbra{\bf0}{\bf0}_{a_A}\otimes
    \ketbra{0}{0}_{a_{\mathcal{G}}}$.
    Conjugating the sign ancilla by Hadamard gates and post-selecting on
    $\mathrm{sgn}=0$ coherently extracts the real part,
    $\frac{1}{2}(e^{i\phi_k\Pi}+e^{-i\phi_k\Pi})$, thereby implementing the
    desired cosine polynomial (see Fig.~\ref{fig:sgn-real-part} for further
    details). Post-selection on $a_{\rm qsvt}=0$ selects the QSVT success
    subspace (see Fig.~\ref{fig:QSVT_Xgrid}). Finally, $\mathrm{PREP}_{w_j}^{\dagger}$ and post-selection of
    the $\rm PREP$ register produce the weighted sum.
    }
    \label{fig:qsvt-cosine-circuit}
\end{figure}

The block encoding $U_A$ is embedded into a Hermitian dilation, yielding the Hermitian operator $U_H = \ketbra{0}{1}\otimes U_A^\dagger + \ketbra{1}{0} \otimes U_A$ (shown in App.~\ref{app:figures}). We then block encode the nodes $\chi$ by classically constructing and injecting them into the phases of controlled-$R_Y$ rotation gates using LCU (see App.~\ref{app:figures}), giving $U_\chi$. Finally, we perform a fixed QSVT sequence, shown in Fig.~\ref{fig:qsvt-cosine-circuit}, that implements Eq.~\eqref{eq:LCCS}.

\subsection{Training and readout}\label{sec:training}

Given a data set $\mathcal{Y} = \{\mathbf{y}_m\}_{m=1}^{n_m}$ drawn from an unknown stationary distribution $p^*$, training constitutes choosing $\boldsymbol\theta$ such that an empirical loss function $\mathcal{L}(\mathcal{Y}, \mathcal{X}, \boldsymbol{\theta})$ is minimized for samples $\mathcal{X} = \{\mathbf{x}_s\}_{s=1}^{n_s}$ obtained by measuring the prepared state $e^{-tH}\ket{\psi}_S$. Throughout the paper we discuss two losses. The first loss is the Kullback--Leibler divergence $\KL(p^*\|p_{\boldsymbol\theta})$~\cite{KLdiv}, which we evaluate on the discretized grid in the emulated experiments, where $p^*$ and $p_{\boldsymbol\theta}$ are both available as vectors. This loss is not sample-accessible on hardware and as such cannot be used by most generative models in practice. As shown in Sec.~\ref{sec:trainability}, however, $\KL$ divergence gradients can be efficiently evaluated by our model from samples. 

The second loss is the squared maximum mean discrepancy ($\MMD^2$), $\MMD^2(P, Q)= \mathbb{E}_{x, x^{\prime} \sim P}\left[k\left(x, x^{\prime}\right)\right] +\mathbb{E}_{y, y^{\prime} \sim Q}\left[k\left(y,y^{\prime}\right)\right]-2 \mathbb{E}_{x \sim P, y \sim Q}[k(x, y)]$ \cite{gretton2012kernel}, a standard choice for implicit quantum generative models~\cite{liu2018differentiable, coyle2020born}.
Minimization is carried out by a classical optimizer acting on $\boldsymbol\theta$. For either loss, each evaluation of the gradient requires one ($\KL$) or two ($\MMD^2$) sample batches (see Sec.~\ref{sec:resources}) and one state preparation per sample.
Details of the numerical experiments are given in App.~\ref{app:training}.


\section{Trainability}\label{sec:trainability}

The optimization problem of Sec.~\ref{sec:training} differs from training a generic parameterized quantum circuit. The optimizer sees $K$ variables regardless of the coarseness of the grid set by $N$ or the data dimensionality $d$. A change of $\boldsymbol\theta$ is passed to the circuit through $O(dr)$ QSVT angles, while the remainder of the compound circuit is set by the discretization and the solver. There is no ansatz-layer count to tune in the usual sense. The only modeling choice is the dictionary $\{v_j\}$, which encodes prior knowledge about the physical system in the same way that a dictionary does in sparse identification of dynamics \cite{brunton2016discovering, boninsegna2018sparse}. Barren plateau arguments for random or highly expressive PQCs \cite{mcclean2018barren, holmes2022connecting} do not apply directly because $\boldsymbol\theta$ does not parameterize a unitary drawn from an approximate design. On the contrary, adding qubits in our construction refines the problem grid and improves the continuum approximation, with the gradient converging to an $n$-independent value. As explained below, assuming close approximation from Eq.~\eqref{eq:LCCS} can be achieved, the problem admits a convex loss landscape owing to the parameterization of the state prepared by the model.

The model in Eq.~\eqref{eq:model_density} corresponds to a subset of the exponential family~\cite{brown1986fundamentals} with natural parameters $\eta_j = -2\theta_j/\sigma^2$ and sufficient statistics $v_j$. Writing $l(\boldsymbol\theta) = \sum_m \log p_{\boldsymbol\theta}(\mathbf{y}_m)$ for the log-likelihood of the data, crucial standard results follow (see App.~\ref{app:expfamily}). First, the stationarity condition $\partial_{\theta_j}l=0$ is equivalent to matching the sufficient statistics,
\begin{equation}\label{eq:moment_matching}
    \mathbb{E}_{p_{\boldsymbol\theta}}[v_j(\mathbf{x})] = \frac{1}{n_m}\sum_{m=1}^{n_m} v_j(\mathbf{y}_m), \qquad j=1,\dots,K,
\end{equation}
which coincides with the stationarity condition of $\KL(p^*\|p_{\boldsymbol\theta})$ and for monomial $v_j$ becomes the standard moment matching. In our implementation, the left-hand side of this equation also becomes the empirical mean from the $n_s$ samples. Second, $-l$ and $\KL(p^*\|p_{\boldsymbol\theta})$ are convex in $\boldsymbol\theta$, and strictly convex whenever the $v_j$ are linearly independent on the support of $p_{\boldsymbol\theta}$. Parameter recovery by maximum likelihood or $\KL$ minimization is therefore a convex problem with a unique optimum, and that optimum is characterized by $K$ moments of the data. On the quantum side this means the circuit need only be sampled enough times to estimate $K$ expectation values $\mathbb{E}_{p_{\boldsymbol\theta}}[v_j]$, rather than to reconstruct $p_{\boldsymbol\theta}$ itself. Using standard results from statistics, $n_s\sim n_m$ balances the statistical errors of the model and data (see App.~\ref{app:expfamily}). We note that score matching~\cite{score_matching} provides a sample-free alternative for training, which could seemingly sidestep the need for the use of quantum hardware during training, but sample-based maximum likelihood estimation has been shown to outperform score matching precisely for the target distributions of interest for our method~\cite{efficiency_score_matching}.
\begin{figure}[t!]
    \centering
    \includegraphics[width=1.0\columnwidth]{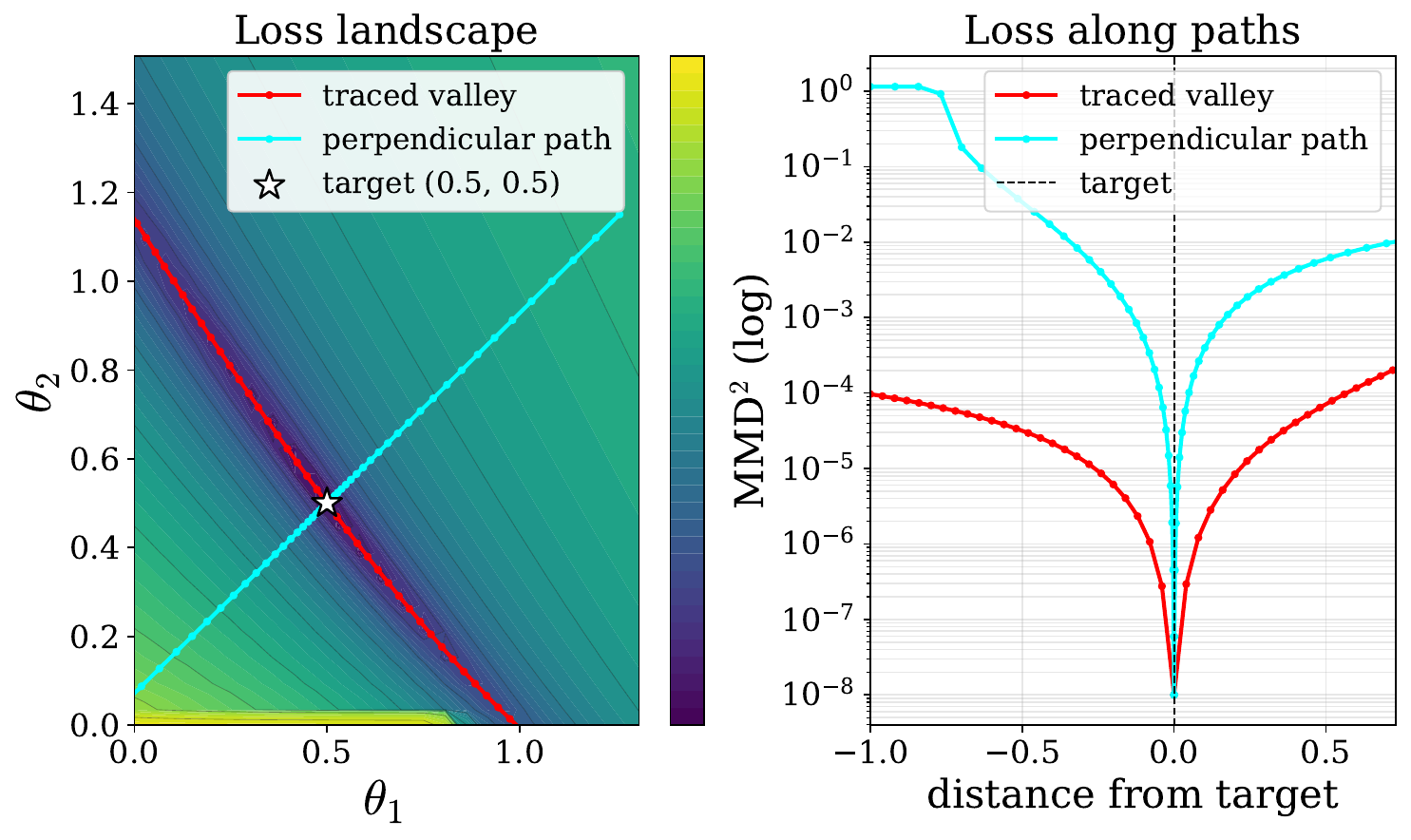}
    \caption{$\MMD^2$ landscape for a 9-qubit one-dimensional quartic model $V(x)=\tfrac{\theta_1}{2}x^2+\tfrac{\theta_2}{4}x^4$ around the true parameters $(\theta_1, \theta_2)=(0.5,0.5)$. Left: contour map with the path along the soft Hessian eigendirection (the valley, red) and the perpendicular stiff direction (cyan). Right: loss along each path as a function of distance from the target. The stiff direction rises by orders of magnitude within a short distance while the valley is shallow, so the landscape is strongly anisotropic. Nevertheless, the loss along the valley still has a well-defined minimum at the target, so the two parameters are not degenerate and the optimum is identifiable.}
    \label{fig:valley_tracer_perp}
\end{figure}

The $\MMD^2$ loss is not in general convex in $\boldsymbol\theta$, even for an exponential family. It is, however, a standard loss used throughout the literature and one that can be estimated directly from samples without the need to estimate moments, which could be useful when the data may lie outside the family spanned by the dictionary $\{v_j\}_{j=1}^K$, when high-degree statistics are needed, or when direct comparability with other models is required. As such, we provide additional training results for the $\MMD^2$ loss in Sec.~\ref{sec:results} and show empirically that it shares the qualitative features of the $\KL$ landscapes for the problems presented in this work. Once again, standard results show that $n_s\sim n_m$ balances the model and data sampling error contributions~\cite{gretton2012kernel}.


\section{Numerical results}\label{sec:results}

Next, we present numerical studies investigating the convergence of the learning algorithm to the target, the trainability behavior with a growing number of potential terms $v_j(\mathbf{x})$, the approximation effects of the quantum solver, and whether the construction extends to correlated, higher-dimensional targets. All results were obtained by classical emulation of the construction of Sec.~\ref{sec:quantum}. Numerical settings are listed in App.~\ref{app:training}.
\begin{figure}[h!]
    \centering
    \includegraphics[width=1.0\columnwidth]{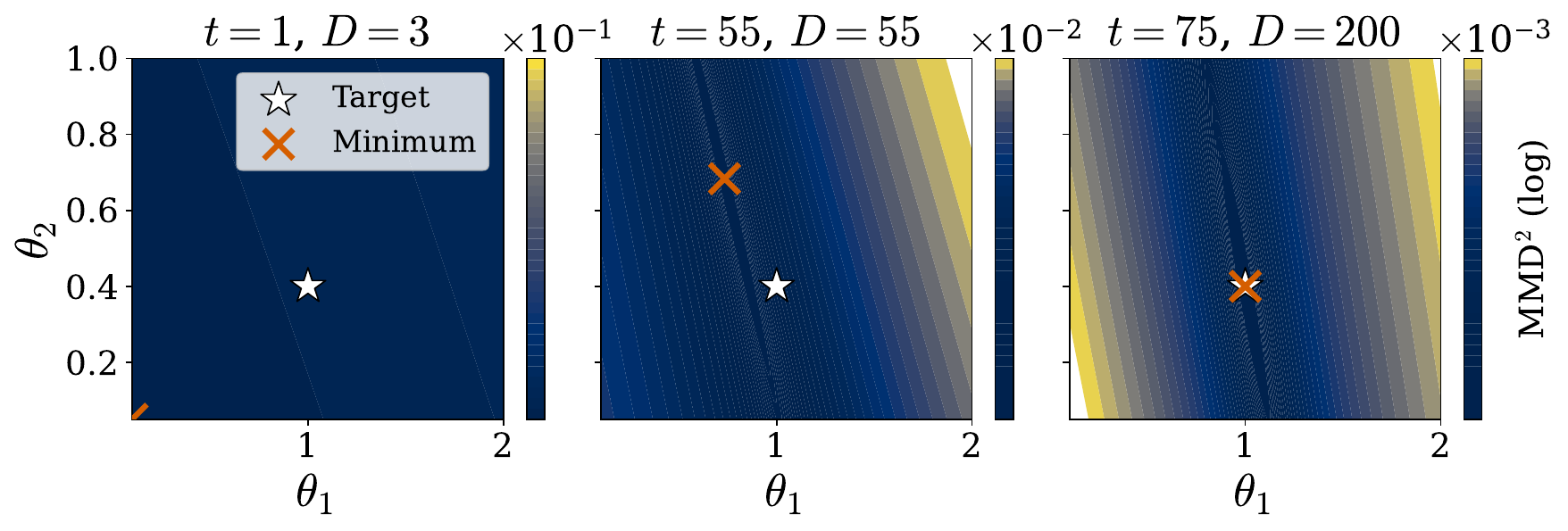}
    \caption{$\MMD^2$ landscapes obtained with the finite-order filter of Eq.~\eqref{eq:LCCS} for the quartic model with $V'(x)=\theta_1 x+\theta_2 x^3$ and true parameters $(\theta_1^*, \theta_2^*)=(1.0,0.4)$, for increasing imaginary time $t$ and quadrature order $D$ from left to right. The red cross marks the minimum found on the grid. At small $t$ or $D$, the filter has not converged and the minimum is displaced. At $t=75$, $D=200$, the landscape and its minimum coincide with the exact one.}
    \label{fig:hsquartic}
\end{figure}

First, we consider a one-dimensional quartic potential $V(x) = \tfrac{\theta_1}{2}x^2+\tfrac{\theta_2}{4}x^4$ with two coefficients $(\theta_1, \theta_2)$ and study the convergence of $\MMD^2$ minimization using the proposed solver. Fig.~\ref{fig:valley_tracer_perp} maps the $\MMD^2$ landscape in the $(\theta_1, \theta_2)$ plane around the target.
The landscape has a single basin whose curvature differs by orders of magnitude between the direction of the largest curvature of the loss at $\boldsymbol{\theta^*}$ and the valley (lowest curvature). Anisotropy of this kind can be explained by the quartic potential, where the two coefficients affect the stationary density in partially compensating ways (a larger quadratic term and a smaller quartic term produce similar widths), so a combination of them is weakly constrained by the data. Importantly, the landscape is not flat. The right-hand panel in Fig.~\ref{fig:valley_tracer_perp} shows that the loss along the valley direction still has a clear minimum at the target, the two coefficients are individually identifiable, and the true stochastic model is recovered. 
\begin{figure}[h!]
    \centering
    \includegraphics[width=\columnwidth]{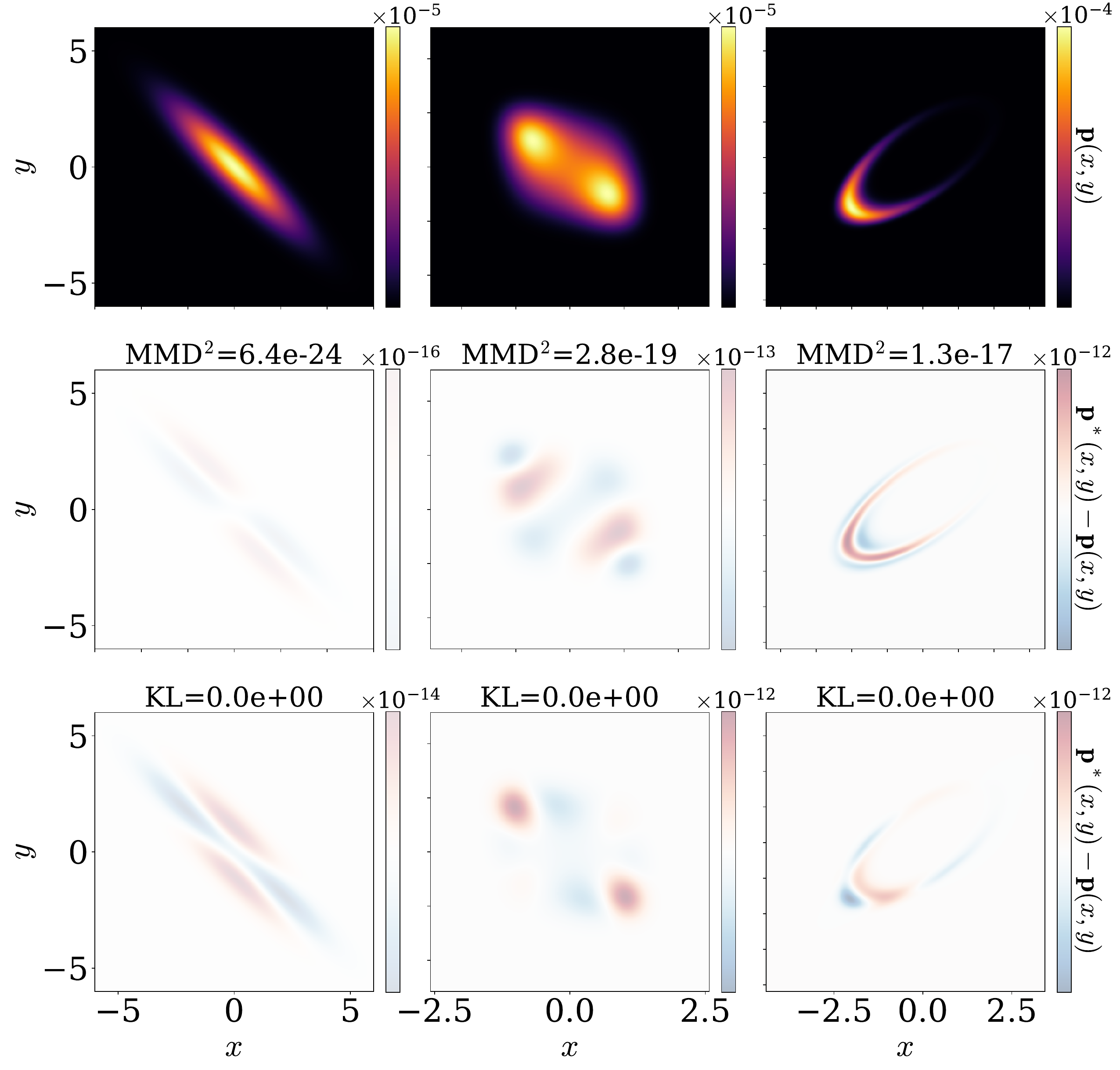}
    \caption{Two-dimensional targets and residuals after training. Columns: an elongated, correlated single-well target (left), a multimodal target with several separated wells (center), and a curved crescent-shaped target (right). Top row: target density $p^*(x,y)$. Middle and bottom rows: residual $p^*-p_{\boldsymbol{ \hat{\theta}}}$ after training with $\MMD^2$ and with $\KL$, respectively, with the final loss. Red corresponds to positive-valued residuals and blue to negative. Residuals are several orders of magnitude below the target amplitude for both losses, so the coupled potential coefficients that generate the correlations are recovered. Training converged within 10--150 iterations for all potentials.}
    \label{fig:2D_potentials}
\end{figure}

The results in Fig.~\ref{fig:valley_tracer_perp} are shown for the exact ground state of the discretized operator. On a quantum device, the state is instead prepared by the finite-order filter of Fig.~\ref{fig:qsvt-cosine-circuit}, raising the question of trainability under finite approximation error. Fig.~\ref{fig:hsquartic} shows the $\MMD^2$ landscape produced by the LCCO-prepared state for increasing imaginary time and quadrature order. For small $t$ the filter has not removed the excited states and the minimum is displaced from the true parameters. For $t$ and $D$ large enough, the landscape converges to the exact one and the minimum coincides with the target. 

\begin{figure}[h!]
    \centering
    \includegraphics[width=\columnwidth]{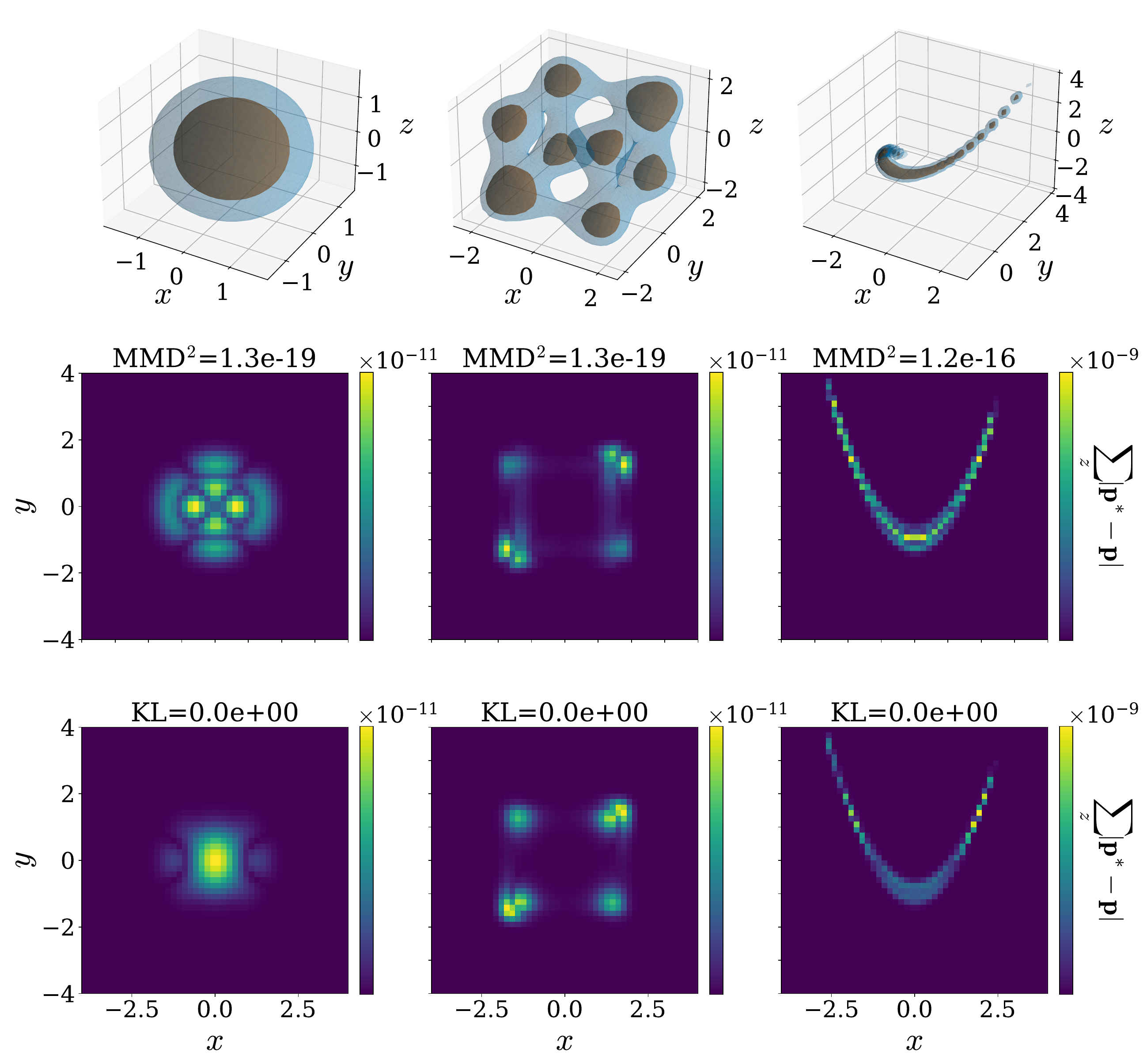}
    \caption{Three-dimensional targets and residuals after training. Columns: a single spherical well (left), a multi-well potential with minima arranged on the vertices of a cube (center), and a helical target (right). Top row: isosurfaces of the target density. Middle and bottom rows: absolute value of the residual $|p^*-p_{\boldsymbol{\hat{\theta}}}|$, summed over $z$, after $\MMD^2$ and $\KL$ training, with the final loss. As in two dimensions, both losses converge to residuals many orders of magnitude below the target amplitude within 40--150 iterations, demonstrating that a few coupled coefficients suffice to define, and to recover, non-separable three-dimensional distributions.}
    \label{fig:3D_potentials}
\end{figure}

As discussed in Sec.~\ref{sec:discussion}, the success of the proposed architecture over classical methods is most promising for correlated multidimensional distributions. Figs.~\ref{fig:2D_potentials} and \ref{fig:3D_potentials} show six such targets, in two and three dimensions, generated by potentials with coupling terms between the coordinates, namely a correlated single well, multi-well potentials with minima on a lattice, and curved targets in which the location of one coordinate's minimum depends on the value of another. In each case the potential is specified by a handful of coefficients, while the discretized distribution has $N^2$ or $N^3$ entries. Training with either $\MMD^2$ or $\KL$ reduces the residual errors to several orders of magnitude below the target amplitude, and the two losses reach optima of comparable quality, with $\KL$ typically requiring fewer iterations. The targets here are small enough to emulate exactly. Extending them to the LCCO-prepared state and to higher $d$ is a natural next step and is discussed in Sec.~\ref{sec:discussion}.

\section{Resource estimation}\label{sec:resources}

The construction of Sec.~\ref{sec:quantum} is a nested chain of fault-tolerant primitives, and a single evaluation of the loss requires it to be executed many times. In this section, we provide a summary of quantum resources required to perform the training algorithm in full on a quantum device, in the simplified case where no cross terms are present, i.e., each $\partial_{x_i}V$ depends only on $x_i$. Cross terms affect only the potential oracle: since $\partial_{x_i}V=\sum_j\theta_j\,\partial_{x_i}v_j$ is linear in $\boldsymbol\theta$ with fixed monomials $\partial_{x_i}v_j$, the diagonal operator $\mathrm{diag}(\partial_{x_i}V)$ can be block encoded as a linear combination of products of QSVT polynomials of position operators, one per coordinate~\cite{gilyen2019quantum}, with $\boldsymbol\theta$ entering through the amplitudes of the combination. A discussion of the extension to correlated potentials, like those shown in Figs.~\ref{fig:2D_potentials} and \ref{fig:3D_potentials}, is included in App.~\ref{app:multidim_corr}, where we argue that modest additional overheads are required. We collect all resource requirements discussed in this section in Table~\ref{tab:resources}. 

During the training, the cost per optimization step is
\begin{equation}\label{eq:train_cost}
    C_{\rm train} = N_{\rm eval}\times \frac{n_s}{P_{\rm succ}} \times C_{\rm prep},
\end{equation}
where $N_{\rm eval}$ is the number of sample batches per step, $n_s$ the number of accepted samples per batch, $P_{\rm succ}$ the post-selection success probability, and $C_{\rm prep}=q\,C_A$ the cost of one attempted state preparation, with the number of queries $q$ and the cost per query $C_A$ given below. The total training cost is $N_{\rm iter}C_{\rm train}$ for $N_{\rm iter}$ optimization steps. 
For both losses, $n_s\sim n_m$ samples balance the model and data sampling errors (Sec.~\ref{sec:trainability}, App.~\ref{app:expfamily}). Because the score of the model is known in closed form, each optimization step needs only a small constant number of batches of samples $N_{\textrm{eval}}=1$ for $\KL$ (App.~\ref{app:expfamily}), and $N_{\textrm{eval}}=2$ for $\MMD^2$, one for each argument of the model-only kernel. Regular QCBM implementations require parameter-shift evaluations, i.e.\ two additional circuit batches per trainable parameter. 
For $\KL$, the convexity of the problem suggests that, at least in the case where a good approximation of the desired stationary state is prepared, the total number of optimization steps is unlikely to dominate the cost.

At each optimization step, a classical optimizer updates $K\le\binom{d+r}{r}-1$ parameters, independent of $N$, and recomputes the $r$ QSVT phases of each potential oracle (one oracle per coordinate), using efficient existing algorithms~\cite{dong2021efficient, Ying_2022}. 
These phases are the only $\boldsymbol\theta$-dependent gates: they enter the block encoding $U_A$ of Fig.~\ref{fig:block-encoding-A} through the QSVT ancilla of the potential oracle $U_{V'}$, which acts on the $n$ qubits of its coordinate.

The circuit width is as follows. The system register holds $dn$ qubits. The position-operator block encoding $U_\mathcal{G} $ uses a PREP register $a_\mathcal{G}$ of $m=\lceil\log_2(n+1)\rceil$ qubitsfor its $n+1$ Pauli terms, and the QSVT that turns it into $U_{V'}$ uses one further ancilla. The LCU of the kinetic and potential terms of $U_A$ uses two ancillas, and for $d>1$ the stacking of the coordinate operators described in App.~\ref{app:multidim_corr} uses an index register of $\lceil\log_2 d\rceil$ qubits. The Hermitian dilation $U_A\rightarrow U_H$ adds one qubit. The cosine QSVT of Fig.~\ref{fig:qsvt-cosine-circuit} adds one ancilla for the quadrature-node encoding, one phase ancilla and one sign ancilla, and for an even number $D$ of quadrature nodes the outer LCU over the $D_+=D/2$ positive nodes adds $\lceil\log_2 D_+\rceil$ qubits for $\mathrm{PREP}_{w_j}$. The total is
\begin{equation}\label{eq:qubits}
    n_{\rm qubits} = dn + \lceil\log_2(n+1)\rceil + \lceil\log_2 d\rceil + \lceil\log_2 D_+\rceil + 7,
\end{equation}
which is logarithmic in the grid size and in the quadrature order. If $V'$ has mixed parity, its even and odd parts are implemented separately and combined by an LCU, which adds one more ancilla.

Having established the required circuit width, we turn to the depth. In one dimension, one query to $U_A$ requires two QFTs on $n$ qubits, $O(n^2)$ gates~\cite{nielsen_chuang_2010qft} (or $O(n\log n)$ approximately~\cite{Nam_2020}), as well as one QSVT of degree $r-1$ on the $O(n)$-gate block encoding of $\mathcal{G}$, i.e.\ $O(rn)$ gates \cite{Kharazi2025explicitblock}, giving $C_A=O(n^2+rn)$ gates per query. 
The subnormalization of $U_A$ is $\alpha_A = \sigma/(\sqrt2\Delta x) + \alpha_{V'}/(\sqrt2\sigma)$. Because $\Delta x=2L/N$, the kinetic contribution grows linearly with the number of grid points, $\alpha_A\propto N$, and this is how the grid resolution enters the depth below. We fix $\alpha_{V'}$ to a bound on $|V'|$ that holds over the whole parameter range explored during training. In this case $\alpha_A$ and the filter in Fig.~\ref{fig:qsvt-cosine-circuit} do not change with $\boldsymbol\theta$.
\begin{table}[t!]
\caption{Resource summary for the training algorithm, for potentials in which each $\partial_{x_i}V$ depends only on $x_i$. $d$: number of coordinates; $n=\log_2N$: qubits per coordinate; $r$: degree of $V$; $K$: number of trainable coefficients; $D$: number of Gauss--Hermite nodes; $\chi_{\max}\approx\sqrt{2D}$: largest node; $t$: imaginary time; $\epsilon$: target precision; $\Delta = E_1-E_0$: spectral gap of $H$; $\zeta$: overlap of the initial state with the ground state; $n_m$, $n_s$: data and accepted model samples. $\bar{\alpha}_A$: multidimensional subnormalization (see App.~\ref{app:multidim_corr}).}
\label{tab:resources}
\begin{ruledtabular}
\begin{tabular}{ll}
Quantity & Scaling\\
\hline
Trainable parameters & $K\le\binom{d+r}{r}-1$\\
Qubits & Eq.~\eqref{eq:qubits}\\
Gates per query to $U_A$, $C_A$ & $O\big(d(n^2+rn)\big)$\\
Subnormalization $\bar\alpha_A$ & $\sqrt d\,\big(\tfrac{\sigma}{\sqrt2\Delta x}+\tfrac{\alpha_{V'}}{\sqrt2\sigma}\big)\propto\sqrt d\,N$\\
Queries per state preparation, $q$ & $O\big(\sqrt t\,\chi_{\max}\bar\alpha_A+\log(1/\epsilon)\big)$\\
$\boldsymbol\theta$-dependent gates per preparation & $O(drq)$\\
Imaginary time & $t=O\big(\Delta^{-1}\log(1/(\epsilon\zeta))\big)$\\
Success probability & $P_{\rm succ}\ge\zeta^2$\\
Accepted samples per step & $n_s\sim n_m$\\
Sample batches per step & $N_{\rm eval}=1$ ($\KL$), $2$ ($\MMD^2$)\\
\end{tabular}
\end{ruledtabular}
\end{table}

The block encoding $U_\chi\otimes U_H$ encodes $\mathrm{diag}(\chi_j/\chi_{\max})\otimes B/\bar\alpha_A$, where $B$ is the Hermitian dilation of $A$ (Sec.~\ref{sec:quantum}), so a single QSVT sequence implements all cosine terms of Eq.~\eqref{eq:LCCS} at once, and additional nodes add qubits to $\mathrm{PREP}_{w_j}$ rather than separate query sequences. In the variable $y\in[-1,1]$ of the block encoding, the largest term is $\cos(2\sqrt t\,\chi_{\max}\bar\alpha_A\,y)$, which the Jacobi--Anger expansion approximates to precision $\epsilon$ by a polynomial of degree~\cite{gilyen2019quantum}
\begin{equation}\label{eq:filter_degree}
    q = O\big(\sqrt t\,\chi_{\max}\bar\alpha_A + \log(1/\epsilon)\big),
\end{equation}
which is the number of queries to $U_A$ per state preparation. Here $\chi_{\max}$ is the largest quadrature node. For a $D$-point quadrature rule, $\chi_{\max}\approx\sqrt{2D}$.

Each query to $U_A$ contains $d$ potential oracles $U_{V'}$, one per coordinate, each carrying $r$ phases, so the $\boldsymbol\theta$-dependent gates appear $O(drq)$ times per state preparation (cf.\ Fig.~\ref{fig:overview}).

The imaginary time itself is $t=O\big(\Delta^{-1}\log(1/(\epsilon\zeta))\big)$, where $\zeta=|\braket{\psi_0}{\psi_{\rm init}}|$ is the overlap of the initial state with the ground state. The gap depends on $\boldsymbol\theta$ and $\sigma$.
All block-encoding ancillas are post-selected on $\ket{0}$. For even $D$ the quadrature weights of the positive nodes sum to $\sqrt{\pi}/2$. The LCU over nodes introduces no additional subnormalization, and the (unnormalized) post-selected state is $e^{-tH}\ket{\psi_{\rm init}}$ up to the filter error. Since $H\succeq0$ with $E_0=0$, the success probability satisfies $P_{\rm succ}\ge\zeta^2$, and fixed-point amplitude amplification reduces the number of repetitions from $O(1/\zeta^2)$ to $O(1/\zeta)$~\cite{gilyen2019quantum}. 

In summary, the circuit width is logarithmic in the grid and in the quadrature order, and the cost of one state preparation, $C_{\rm prep}=qC_A$, is polynomial in $N$, $d$, $t$, $D$ and $\log(1/\epsilon)$ rather than in the $N^d$ grid size. We note that some quantities can still grow exponentially as the number of coordinates grows, namely the inverse gap $\Delta^{-1}$ (potentially in the small-noise multi-well regime) and the inverse overlap $\zeta^{-1}$. Identifying the precise regime of advantage, as well as the classical sampling limits, sets a goal for future work.


\section{Discussion and conclusions}\label{sec:discussion}

We have proposed a quantum generative model in which the trainable variables are coefficients of the potential of a stochastic differential equation and the circuit is a quantum solver for the associated Fokker--Planck problem. The construction avoids the usual pitfalls of scaling variational quantum machine learning models: the circuit is fixed and algorithmically structured, and the model is parameterized at the level of the differential operator it block encodes. The number of trainable parameters is set by the physics, is independent of the grid, and can remain small while the prepared state occupies an exponentially large configuration space. The optimization problem is closer to model discovery than to distribution fitting: the optimum is a drift field, likelihood-based training is convex in the natural parameters because the stationary state is an exponential family, and the optimum is characterized by $K$ moments of the data. The approximations of the solver (discretization, finite imaginary time, finite quadrature order) are controllable, with the learned model converging to the exact one as they are refined. The numerical results support each of these points in one dimension and, with coupled potentials, in two and three dimensions, within classical emulation.

The regime in which the architecture is most attractive is that of many coupled stochastic degrees of freedom, for which a $d$-dimensional grid with $N$ points per axis is out of reach for classical grid-based methods but occupies $d\log_2 N$ qubits. The two- and three-dimensional examples of Sec.~\ref{sec:results} already use coupling terms, and the block encoding construction does not incur qubit or gate costs exponential in the problem dimension. Couplings of this kind are well established on the classical side. Quartic couplings $x_i^2x_j^2$ and Ginzburg--Landau potentials produce multimodal, non-Gaussian correlated states in coarse-grained molecular~\cite{boninsegna2018sparse} and field-theoretic models~\cite{landaumodel}. More generally, non-Gaussian pairwise Markov random fields have the same polynomial exponential-family form, and identifying their non-zero couplings is a central task in statistical learning~\cite{graphical_models}.

A PQC-based quantum circuit Born machine trained with the same data would return a set of gate angles with no direct physical meaning. Here the target parameters $\boldsymbol\theta^*$ make up the coefficient vector of a drift field, and the associated overdamped Langevin equation with $V(\cdot;\boldsymbol{\hat{\theta}})$ is a stochastic model of the data that can be simulated forward, perturbed, or inspected. Identifiability is inherited from the exponential-family structure: if the data-generating potential lies in the span of the dictionary and $\sigma$ is known, the maximum-likelihood estimator $\boldsymbol{\hat{\theta}}$ converges to the true coefficients $\boldsymbol\theta^*$ as $n_m\to\infty$. If the dictionary is misspecified, $\boldsymbol{\hat{\theta}}$ is the moment projection of $p^*$ onto the model family, which is still the best stochastic model of the given form.

Regarding classical simulation hardness, the Hilbert space dimension alone does not establish a quantum advantage. Identifying the boundary of classical simulation requires an end-to-end comparison with state-of-the-art classical sampling methods for the same family of distributions. 
On the quantum side, this comparison must account for the state-preparation cost of Sec.~\ref{sec:resources}, including the gap dependence of $t$ and the post-selection success probability (App.~\ref{app:errors}). It needs to include the sample complexity of the loss and the number of optimization steps. On the classical side, the relevant benchmark is the best available sampler for the same family. In the small-noise, multi-well regime, mixing can be the main bottleneck, rather than grid size \cite{beck2016rare, giles2015multilevel}. 
The present work provides a model architecture for this comparison: a physics-informed generative model with a low-dimensional trainable parameter space and a structured state-preparation circuit inherited from quantum differential equation solvers. The framework hence opens a route to identifying regimes in which quantum state preparation offers an advantage over classical sampling.

\begin{acknowledgments}
The authors acknowledge discussions with the Haiqu team on the subject of differential equation solvers. M.\,J. and O.\,K. acknowledge support from the UK EPSRC award under Agreement No.~EP/Z53318X/1 (QCi3 Hub). O.\,K. acknowledges support from the UK EPSRC Open Plus Fellowship award ``Quantum Generative Modelling: from Foundations to Applications''.
\end{acknowledgments}

\textit{AI statement.} The original idea was proposed and developed by humans. AI tools were used to aid with coding and writing parts of the manuscript. The authors take responsibility for results and presentation.

\bibliography{references}

\appendix

\section{Circuit considerations for correlated multidimensional potentials}
\label{app:multidim_corr}

The main text constructs the circuit for a single coordinate and, in
Sec.~\ref{sec:resources}, for $d$ coordinates when each $\partial_{x_i}V$ depends on $x_i$ only. Here we describe what changes when the potential couples the coordinates, as for the targets of Figs.~\ref{fig:2D_potentials} and~\ref{fig:3D_potentials}, and as is required for constructing problems of interest. In $d$ dimensions the operator to be filtered is $H=\sum_{i=1}^{d}A_i^\dagger A_i$, with $A_i=\frac{\sigma}{\sqrt2}\partial_{x_i}+\frac{1}{\sqrt2\,\sigma}\partial_{x_i}V$. Each $A_i$ is block encoded as $U_A$ in Fig.~\ref{fig:block-encoding-A}, with the shift operators acting on the register of $x_i$, and only the potential oracle changes: for each $i$ we need a block encoding $U_{\partial_iV}$ of the diagonal operator $\mathrm{diag}\big(\partial_{x_i}V(\mathbf{x})\big)/\alpha_{\partial_iV}$ on the full grid. For a separable potential $U_{\partial_iV}$ is the circuit $U_{V'}$ of Fig.~\ref{fig:QSVT_Xgrid} acting on register $i$. For a correlated potential it acts on several registers. The kinetic terms, the Hermitian dilation (Fig.~\ref{fig:dilation}), the filter (Fig.~\ref{fig:qsvt-cosine-circuit}, up to its projector, see below) and the exponential-family structure of App.~\ref{app:expfamily} are unchanged, since $V(\mathbf{x};\boldsymbol\theta)$ remains linear in $\boldsymbol\theta$. 

With the monomial dictionary of Sec.~\ref{sec:model}, each component of the gradient is a polynomial $\partial_{x_i}V(\mathbf{x};\boldsymbol\theta)=\sum_{t=1}^{T_i} c_t(\boldsymbol\theta)\prod_{k=1}^{d}x_k^{a_{t,k}}$, whose coefficients $c_t$ are linear in $\boldsymbol\theta$ and whose exponents $\boldsymbol a_t$ are fixed by the dictionary.
On the grid we write $x_k=\alpha_{\mathcal G}\,y_k$, where $\alpha_{\mathcal G}=\max_j|x_j|$ is the subnormalization of $U_{\mathcal G}$, such that $y_k\in[-1,1]$. Each monomial is then a product of univariate factors $y_k^{a}$, and each factor is a fixed circuit on register $k$ with the polynomial $P(y)=y^{a}$. Factors acting on different registers are applied side by side, each with its own ancillas, and post-selecting all of these ancillas on $|0\rangle$ block encodes the product of the factors. The $T_i$ monomials are then combined by an LCU with weights $w_t=|c_t|\,\alpha_{\mathcal G}^{|\boldsymbol a_t|}$, $|\boldsymbol a_t|=\sum_k a_{t,k}$, on a selector register of $\lceil\log_2T_i\rceil$ qubits, with the signs of $c_t$ applied as phases on the selector. 
Since the branches are mutually exclusive, the factor ancillas are shared between terms: the oracle needs $k_{\max}$ PREP registers of $U_{\mathcal G}$ and $k_{\max}$ QSVT ancillas, where $k_{\max}\le d$ is the largest number of coordinates in a single monomial.

The parameters enter the circuit differently from the construction described in the main text. The QSVT phases of the factors are fixed for the whole training run, and $\boldsymbol\theta$ enters only through the amplitudes of the selector state preparation, which follow in closed form from the weights $w_t$. The same construction could also be used in one dimension.

Other routes to multivariate diagonal block encodings exist, for instance an LCU of Fourier components~\cite{rosenkranz2025quantumstate} or Chebyshev feature maps~\cite{kyriienko_piqml}. The construction above has the advantage of reusing $U_{\mathcal G}$ and the QSVT circuit of Fig.~\ref{fig:QSVT_Xgrid} without modification.

The $d$ operators $A_i$ are combined into $A=\sum_{i=1}^{d}|i\rangle\langle0|_{a_{\rm idx}}\otimes A_i$ on an index register $a_{\rm idx}$ of $\lceil\log_2d\rceil$ qubits, so that $A^\dagger A=|0\rangle\langle0|_{a_{\rm idx}}\otimes H$. A state preparation $\mathrm{PREP}_{\rm idx}$ of the uniform superposition over the $d$ index states, followed by a SELECT over the $U_{A_i}$, with $a_{\rm idx}$ retained rather than uncomputed, block encodes $A/\bar\alpha_A$ with $\bar\alpha_A=\sqrt d\,\alpha_A$.
All $U_{A_i}$ share the same subnormalization $\alpha_A$, fixed in advance as a bound over the coordinates and the training range. One query to the stacked operator costs one query to each $U_{A_i}$. 

This block encoding is rectangular: its input is restricted to $a_{\rm idx}=0$, while its output is not. 
A single flag qubit $a_{\rm f}$ that records whether $a_{\rm idx}\neq0$ before $\mathrm{PREP}_{\rm idx}$ is applied (a CNOT for $d=2$, a multi-controlled NOT in general) resolves this. The modified oracle $U_A'=\mathrm{SELECT}\,(\mathrm{PREP}_{\rm idx}\otimes I)\,\mathrm{C}_{a_{\rm idx}\to a_{\rm f}}$ satisfies $\langle0|_{a_{\rm f}}U_A'|0\rangle_{a_{\rm f}}=A/\bar\alpha_A$ as a square operator on $a_{\rm idx}$ and the system, with all other ancillas of the $U_{A_i}$ projected on $|0\rangle$. The filter of Fig.~\ref{fig:qsvt-cosine-circuit} therefore applies unchanged, with $a_{\rm f}$ added to $\Pi$ and $a_{\rm idx}$ treated as part of the system.

Relative to Eq.~\eqref{eq:qubits}, the correlated oracle adds $(k_{\max}-1)(m_{\mathcal G}+1)+\lceil\log_2T\rceil$ qubits, with $T=\max_iT_i$, $m_{\mathcal G}$ the size of the PREP register of $U_{\mathcal G}$ ($\lceil\log_2(n+1)\rceil$ in Eq.~\eqref{eq:qubits}). The number of terms limits the approach: a dense polynomial $\partial_{x_i}V$ of degree $r-1$ in $d$ variables has up to $\binom{d+r-1}{r-1}$ monomials, so the construction is suited to sparse couplings. The main cost of correlations is therefore the normalization $\alpha_{\partial_iV}$, which enters $q$ through $\bar\alpha_A$, rather than the circuit width.

\section{Hermitianization of the Fokker--Planck operator}\label{app:hermitianization}

We write Eq.~\eqref{eq:fp} as $\partial_t p = \mathcal{L}p$ with $\mathcal{L}p = \nabla\cdot(p\nabla V) + \tfrac{\sigma^2}{2}\Delta p$, and substitute $p = e^{-V/\sigma^2}\psi$. Using $\nabla p = e^{-V/\sigma^2}(\nabla\psi - \psi\nabla V/\sigma^2)$ and collecting terms, the first-order derivatives of $\psi$ cancel and one obtains $\partial_t\psi = -H\psi$ with $H$ as in Sec.~\ref{sec:quantum}. The same $H$ is obtained by the positive semidefinite factorization
\begin{equation}\label{eq:psdf}
\begin{aligned}
    H &= \sum_{i=1}^{d} A_i^\dagger A_i,\\
    A_i &= \frac{\sigma}{\sqrt2}\partial_{x_i} + \frac{\partial_{x_i}V}{\sqrt2\,\sigma},\quad
    A_i^\dagger = -\frac{\sigma}{\sqrt2}\partial_{x_i} + \frac{\partial_{x_i}V}{\sqrt2\,\sigma}, 
\end{aligned}
\end{equation}
since $A_i^\dagger A_i = -\tfrac{\sigma^2}{2}\partial_{x_i}^2 + \tfrac{(\partial_{x_i}V)^2}{2\sigma^2} - \tfrac{1}{2}\partial_{x_i}^2V$, the cross terms $\pm\tfrac12[\partial_{x_i},\partial_{x_i}V]$ combining into $-\tfrac12\partial_{x_i}^2V$. This is a supersymmetric form of the Fokker--Planck operator \cite{bernstein1984susy, risken1989}. $H\succeq0$, with $A_i\psi_0 = 0$ for all $i$ when $\psi_0\propto e^{-V/\sigma^2}$, so the ground energy is exactly zero and the ground state is the square root of the stationary density. Note that we block encode the first-order operators $A_i$ and not $H$ directly, since they contain $\partial_{x_i}V$ (degree $r-1$) rather than $(\partial_{x_i}V)^2$ (degree $2r-2$).


\section{Approximation errors}\label{app:errors}

The distance between the distribution sampled from the circuit and the exact stationary density $p_{\rm ss}$ is bounded by the sum of the following contributions, listed in the order in which they enter the construction. The discretization terms dominate in the present experiments and we discuss them first.

\paragraph{Grid discretization.} The central-difference operator has truncation error $O(\Delta x^2)$ in $\partial_x$, and the discretized ground state differs from the sampled continuum $\sqrt{p_{\rm ss}}$ by $O(\Delta x^2\,\|\partial_x^3\psi_0\|)$.

\paragraph{Finite-domain truncation and boundary conditions.} Restricting to $[-L,L]$ neglects the stationary mass outside the interval, $\int_{\rm outside}p_{\rm ss}$, and the periodic (QFT) boundary couples the two ends. For a confining polynomial potential both are exponentially small in the domain half-width $L$ measured in units of the width of $p_{\rm ss}$, but they are independent of $\Delta x$ and can set the floor if $L$ is too small.

\paragraph{Block-encoding subnormalization.} $U_A$ encodes $A/\alpha_A$ with $\alpha_A=\sigma/(\sqrt2\Delta x)+\alpha_{V'}/(\sqrt2\sigma)$. This is exact but rescales the spectrum, so the imaginary time entering the circuit is $t\alpha_A^2$ and the required $t$ and $D$ scale accordingly. 

\paragraph{Polynomial (QSVT) approximation.} Each cosine and the potential polynomial are implemented to precision $\epsilon_{\rm poly}$. The error in the encoded block is at most $\epsilon_{\rm poly}$ per query and accumulates additively over the $O(\deg)$ queries. In the emulations of Sec.~\ref{sec:results} the polynomials are exact and this term is zero. 

\paragraph{Quadrature order.}  The error is that of the $D$-point quadrature rule applied to $\cos(2\sqrt t\lambda \chi)$ for each eigenvalue $\lambda$ of $H$ and is non-monotone in $D$.

\paragraph{Finite imaginary time.} 
After time $t$ the unnormalized state is $\sum_k c_k e^{-tE_k}\ket{E_k}$, so after normalization the total variation distance (TVD) to the ground-state distribution decays as $e^{-t\Delta}$, with a prefactor inversely proportional to the initial overlap $\zeta$. The required $t$ is set by the gap $\Delta$ of $H$.

\paragraph{Sampling and post-selection.} The loss is estimated from $n_s$ circuit samples and $n_m$ data samples. The statistical error of $\MMD^2$ is $O(1/\sqrt{\min(n_s,n_m)})$ \cite{gretton2012kernel}, and unsuccessful post-selection reduces the effective $n_s$ by the success probability. An attempt succeeds when all ancillas are measured in $\ket{0}$. The subnormalizations $\bar{\alpha}_A$ (see App.~\ref{app:multidim_corr}) and $\chi_{\textrm{max}}$ are absorbed into the argument of the QSVT polynomial, $f(y) = \cos(2\sqrt{t} \chi_{\textrm{max}} \bar{\alpha}_A y)$, so they increase the number of queries $q$ in Eq.~\eqref{eq:filter_degree} but cause no post-selection loss. The post-selected state is $f(H)\ket{\psi_{\textrm{init}}} \approx e^{-tH}\ket{\psi_{\textrm{init}}}$. Hence 
\begin{equation}
    P_{\textrm{succ}} = \sum_k |c_k|^2 e^{-2tE_k} + O(\epsilon_{\textrm{filt}}) \geq \zeta^2 e^{-2tE_0} + O(\epsilon_{\textrm{filt}}),
\end{equation}
with $c_k = \langle E_k|\psi_{\textrm{init}}\rangle$ and $\zeta = |c_0|$. Here $E_0=0$ for the continuum operator; on the grid $H=A^\dagger A$ remains positive semidefinite and $E_0\ge0$ arises only from the finite domain and the derivative approximation, and $\epsilon_{\textrm{filt}}= \|f(H)\ket{\psi_{\textrm{init}}} - e^{-tH}\ket{\psi_{\textrm{init}}}\|$ is the combined quadrature and polynomial error described above.


\section{Training details}\label{app:training}

In Fig.~\ref{fig:valley_tracer_perp}, the target and trial distributions are evaluated analytically on a 512-point (9-qubit) grid. The loss function $\mathcal{L}$ is then evaluated using the standard radial basis function (RBF) kernel $K_{ij}=\exp\left[-(x_i-x_j)^2/2h^2\right]$, with $h=0.6$, such that $\mathcal{L}=\sum_{i=1}^{512}\sum_{j=1}^{512}(p_i^*-q_i)K_{ij}(p_j^*-q_j)$. Here, $p^*$ denotes the target distribution parameterized by the monomial coefficients $(\theta_1^*, \theta_2^*)=(0.5,0.5)$ (see Table~\ref{tab:monomial_coefficients}). Exact probabilities are used throughout, so shot noise is absent.

In Fig.~\ref{fig:hsquartic}, the positive-semidefinite Hamiltonian $H=A^{\dagger}A$ is constructed on a 256-point (8-qubit) periodic grid. The operator $A$ is defined as $A=\frac{\sigma}{\sqrt{2}}\mathcal{D}+\frac{1}{\sqrt{2}\sigma}\left(\theta_1 x+\theta_2 x^3\right)$, where $\mathcal{D}=F^{\dagger}\operatorname{diag}(ik)F$ is the periodic spectral derivative operator and $F$ is the discrete Fourier transform. The Hamiltonian is diagonalized as $H=W\Lambda W^{\dagger}$, giving the singular values of $A$ as $\sigma_i(A)=\sqrt{\lambda_i(A^{\dagger}A)}$. The finite-order Hubbard--Stratonovich filter $f_{t,D}$ is then applied to this singular-value spectrum (details of constructing $f_{t,D}$ are given in Sec.~\ref{sec:quantum}), approximating imaginary-time evolution as $\ket{\psi_{t,D}}=Wf_{t,D}(\sqrt{\Lambda})W^{\dagger}\ket{\psi_0}$. The initial state is chosen such that $\ket{\psi_0}\propto\sum_i\psi_0(x_i)\ket{x_i}$, with $|\psi_0(x)|^2\sim\mathcal{N}(0,0.2^2)$. The resulting 256-component probability distribution $q_i=\left|\left\langle x_i\mid\psi_{t,D}\right\rangle\right|^2$ is then compared with the exact 256-component target distribution using $\MMD^2$ with the RBF kernel bandwidth $h=0.6$, producing one scalar loss for each $(\theta_1, \theta_2)$. The dominant approximation error arises from the finite-order Hubbard--Stratonovich representation, since the oscillatory terms $\cos(2 \sqrt{t} \chi \sigma_{\text{max}}(A))$ become increasingly difficult to resolve as the evolution time, domain range, or spectral range increases, requiring a correspondingly larger quadrature order $D$.

In Figs.~\ref{fig:2D_potentials} and \ref{fig:3D_potentials}, analytical target distributions are fitted using $\MMD^2$ and $\KL$ divergence. Each potential is evaluated on a uniform $1024\times1024 \ (\times1024)$ grid using finite-difference discretization over $[-6,6]^2$, with $\sigma=1$. 
$\MMD^2$ uses a Gaussian bandwidth $h=0.6$ for $d=2$ and $h=0.4$ for $d=3$ distributions. Optimization is performed using L-BFGS-B with analytical gradients and a history of 20 correction pairs.
To preserve the required structure of the $d=2$ potentials during optimization, constrained coefficients are reparameterized using unconstrained variables. For the elliptical potential, a Cholesky parameterization enforces $a>0$, $b>0$, and $c^2<4ab$ for the coefficients $(a,b,c)$ of $(x^2,y^2,xy)$. For the bimodal potential, the quartic coefficients are parameterized as
$\lambda_x=e^{2\eta_p}, \lambda_y=e^{2\eta_q}, \kappa=\eta_s^2-2e^{\eta_p+\eta_q}$, which guarantees a non-negative quartic contribution. The $d=2$ crescent and $d=3$ potentials do not require reparameterization.
\begin{table}[t]
\centering

\resizebox{\columnwidth}{!}{%
\begin{tabular}{|c|c|}
\hline
\textbf{Coefficient vector} & \textbf{Monomial dictionary} \\
\hline

\multicolumn{2}{|c|}{\textbf{1D}} \\
\hline

$(0.50, 0.50)$ & $\big((2),\ (4)\big)$ \\
\hline

\multicolumn{2}{|c|}{\textbf{2D}} \\
\hline

$(0.50,\ 0.50,\ 0.90)$
&
$\big((2,0),\ (0,2),\ (1,1)\big)$
\\[4pt]
\hline

$\begin{aligned}
(&-0.20,\ 0.30,\ 0.90,\\
 &0.50,\ 0.50,\ 0.20)
\end{aligned}$
&
$\begin{aligned}
\big(&(2,0),\ (0,2),\ (1,1),\\
     &(4,0),\ (0,4),\ (2,2)\big)
\end{aligned}$
\\[5pt]
\hline

$\begin{aligned}
(&6.00,\ 0.37,\ 0.26,\ -5.35,\\
 &9.32,\ -8.74,\ 1.19,\ -4.15,\\
 &7.51,\ -6.79,\ 3.18)
\end{aligned}$
&
$\begin{aligned}
\big(&(0,0),\ (1,0),\ (0,1),\ (2,0),\\
     &(1,1),\ (0,2),\ (4,0),\ (3,1),\\
     &(2,2),\ (1,3),\ (0,4)\big)
\end{aligned}$
\\[5pt]
\hline

\multicolumn{2}{|c|}{\textbf{3D}} \\
\hline

$\begin{aligned}
(&0.50,\ 0.60,\ 0.40,\ 0.08,\ 0.10,\\
 &0.07,\ 0.15,\ 0.10,\ 0.12)
\end{aligned}$
&
$\begin{aligned}
\big(&(2,0,0),\ (0,2,0),\ (0,0,2),\\
     &(4,0,0),\ (0,4,0),\ (0,0,4),\\
     &(2,2,0),\ (2,0,2),\ (0,2,2)\big)
\end{aligned}$
\\[5pt]
\hline

$\begin{aligned}
(&4.65,\ -1.80,\ -1.52,\ -1.37,\\
 &-0.10,\ -0.08,\ -0.12,\\
 &0.40,\ 0.45,\ 0.35)
\end{aligned}$
&
$\begin{aligned}
\big(&(0,0,0),\ (2,0,0),\ (0,2,0),\\
     &(0,0,2),\ (1,1,0),\ (1,0,1),\\
     &(0,1,1),\ (4,0,0),\\
     &(0,4,0),\ (0,0,4)\big)
\end{aligned}$
\\[5pt]
\hline

$\begin{aligned}
(&15.0,\ 24.0,\ -12.0,\ -15.5,\\
 &12.0,\ 12.0,\ 3.0,\ -15.6,\\
 &5.07,\ -6.00,\ 0.75)
\end{aligned}$
&
$\begin{aligned}
\big(&(0,0,0),\ (0,1,0),\ (0,0,1),\\
     &(2,0,0),\ (0,2,0),\ (0,0,2),\\
     &(3,0,0),\ (2,1,0),\\
     &(4,0,0),\ (3,0,1),\ (6,0,0)\big)
\end{aligned}$
\\

\hline
\end{tabular}%
}
\caption{Coefficient vectors and corresponding monomial dictionaries for Figs.~\ref{fig:valley_tracer_perp}, \ref{fig:2D_potentials}, and \ref{fig:3D_potentials}. 
Each monomial is represented by its exponent tuple. For example, 
$(1,2)$ represents $xy^2$ and $(1,2,3)$ represents $xy^2z^3$.}
\label{tab:monomial_coefficients}
\end{table}


\section{Exponential-family properties of the model}\label{app:expfamily}

The model density of Eq.~\eqref{eq:model_density} has the exponential-family form $p_{\boldsymbol\theta}(\mathbf{x})=\exp\!\big(\sum_j\eta_j v_j(\mathbf{x})-\log Z(\boldsymbol\theta)\big)$ with natural parameters $\eta_j=-2\theta_j/\sigma^2$ and sufficient statistics $v_j$ \cite{brown1986fundamentals}. We record the properties stated in Sec.~\ref{sec:trainability}.

For a sample $\{\mathbf{y}_m\}_{m=1}^{n_m}$ the log-likelihood is $l(\boldsymbol\theta)=-\tfrac{2}{\sigma^2}\sum_j\theta_j\sum_mv_j(\mathbf{y}_m)-n_m\log Z(\boldsymbol\theta)$. The maximum likelihood estimator (MLE) corresponds to $\nabla_{\boldsymbol{\theta}}l=0$. Differentiating $\log Z(\boldsymbol\theta)$, with $Z(\boldsymbol{\theta})=\int e^{-(2/\sigma^2)\sum_a\theta_av_a}\,d\mathbf{x}$, gives
\begin{equation}
    \partial_{\theta_j}\log Z(\boldsymbol\theta)
    = -\frac{2}{\sigma^2}\,\mathbb{E}_{p_{\boldsymbol\theta}}[v_j(\mathbf{x})].
    \label{eq:diflogz}
\end{equation}
Thus, $\partial_{\theta_j}l = 0$ gives exactly Eq.~\eqref{eq:moment_matching}. 

Likewise $\KL(p^*\|p_{\boldsymbol\theta})= \int p^*\log{\frac{p^*}{p_{\boldsymbol{\theta}}}}\, \mathrm{d}\mathbf{x} =\int p^*\log p^*\,\mathrm{d}\mathbf{x} - \sum_a\eta_a\mathbb{E}_{p^*}[v_a]+\log Z(\boldsymbol\theta)$. Setting $\partial_{\theta_j}\KL(p^*\|p_{\boldsymbol\theta}) = 0$ and utilizing Eq.~\eqref{eq:diflogz} again, we arrive at Eq.~\eqref{eq:moment_matching} with the empirical average replaced by $\mathbb{E}_{p^*}[v_j]$. Thus we show that minimizing $\KL$, maximizing likelihood and matching the $K$ moments of the sufficient statistics are the same problem.

The convexity of the log-likelihood (and, equivalently, $\KL$ loss via moment-matching) is shown below. 
The only term of $-l$ that is nonlinear in $\boldsymbol\theta$ (and thus could make it non-convex) is $n_m\log Z(\boldsymbol\theta)$. Differentiating Eq.~\eqref{eq:diflogz} once more to obtain entries of the Hessian gives,
\begin{align}
    \partial_{\theta_k}\partial_{\theta_j}\log Z(\boldsymbol\theta)
    &= \frac{4}{\sigma^4}\Big(\mathbb{E}_{p_{\boldsymbol\theta}}[v_jv_k]-\mathbb{E}_{p_{\boldsymbol\theta}}[v_j]\,\mathbb{E}_{p_{\boldsymbol\theta}}[v_k]\Big) \nonumber\\
    &= \frac{4}{\sigma^4}\,\mathrm{Cov}_{p_{\boldsymbol\theta}}[v_j,v_k],
\end{align}
a covariance matrix, which is positive semidefinite. A positive semidefinite Hessian means $\log Z$ is convex in $\boldsymbol\theta$ and as such $-l$ and $\KL(p^*\|p_{\boldsymbol\theta})$ are convex as well. 

As mentioned in the main text, although $\KL$ is an explicit loss, training by moment matching requires only sample access to the data and to the model. Reconstructing the full probability mass function from model samples, as an explicit loss would generally require, means estimating the probability of each of the $N^d$ grid points. Moment matching instead requires only the $K$ moments $\mathbb{E}[v_j]$ of the data and of the model. For independent samples, the empirical mean of $v_j$ has standard error $\sqrt{\mathrm{Var}[v_j]/n_m}$ for the $n_m$ data samples and $\sqrt{\mathrm{Var}[v_j]/n_s}$ for the $n_s$ model samples, where near convergence $\mathrm{Var}[v_j]$ is the same for both. Since the two estimates are independent, their errors add in quadrature, and the error in matching them is $\sqrt{\mathrm{Var}[v_j]\,(1/n_m+1/n_s)}$. The data term is fixed, so taking $n_s\gtrsim n_m$ gives diminishing returns.

\makeatletter
\setlength{\@fptop}{0pt}
\setlength{\@fpsep}{10pt}
\setlength{\@fpbot}{0pt plus 1fil}
\makeatother

\section{Additional circuit diagrams}\label{app:figures}

The following figures give the remaining circuit-level building blocks referenced throughout the main text and previous appendices. The quadrature-node encoding circuit and the potential-oracle circuits mentioned in Sec.~\ref{sec:quantum} are shown in Figs.~\ref{fig:UX} and~\ref{fig:QSVT_Xgrid}. Hermitian dilation is shown in Fig.~\ref{fig:dilation} and the real-part extraction gadget in Fig.~\ref{fig:sgn-real-part}. Additionally, we show the PREP (Fig.~\ref{fig:PREP_Xgrid}) and position-operator (Fig.~\ref{fig:SEL_Xgrid}) circuits.

The quadrature-node encoding circuit and the potential-oracle circuits mentioned in Sec.~\ref{sec:quantum} are shown in Figs.~\ref{fig:UX} 

\begin{appfigure}
\centering
\(\displaystyle
U_\chi=\sum_j \ket{j}\!\bra{j}_{\mathrm{PREP}_{w_j}}\otimes R_Y(\phi_j)
\)\par\smallskip
\begin{adjustbox}{max width=\columnwidth,center}
\begin{quantikz}[row sep=0.25cm, column sep=0.3cm]
\gate{\mathrm{PREP}_{w_j}}
& \muxctrl{1}
& \cdots
& \muxctrl{1}
& \\
\lstick{$\ket{0}_{a_\chi}$}
& \gate{R_Y(\phi_0)}
& \cdots
& \gate{R_Y(\phi_{D_+-1})}
& \meter{0}
\end{quantikz}
\end{adjustbox}
\caption{The unitary construction $U_\chi$ of the quadrature nodes $\chi_j$ using $R_Y$ gates with angles $\phi_j=2\arccos\left(\frac{\chi_j}{\chi_{\max}}\right)$ such that $R_Y(\phi_j)\ket{0}_{a_\chi} = \frac{\chi_j}{\chi_{\rm max}}\ket{0}_{a_\chi} + \sqrt{1- \left(\frac{\chi_j}{\chi_{\rm max}}\right)^2}\ket{1}_{a_\chi}$.
Post-selecting on $a_\chi=0$ projects onto the correctly encoded, normalized quadrature nodes. The signal operator is then the joint block encoding $U_S = U_\chi \otimes U_H$.}
\label{fig:UX}
\end{appfigure}

\begin{appfigure}
\centering
\(\displaystyle
U_\mathcal{G}\;\xrightarrow{\;\mathrm{QSVT}\;}\;U_{V'(\mathcal{G})}
\)\par\smallskip
\begin{adjustbox}{max width=\columnwidth,center}
$\left.\vcenter{\hbox{%
\begin{quantikz}[row sep=0.3cm,column sep=0.25cm]
\lstick{$\ket{0}_{a_\mathrm{qsvt}}$}
& \gate{H} & \targ{} & \gate{R_Z(\phi_0)} & \targ{} & \qw
& \targ{} & \gate{R_Z(\phi_1)} & \targ{} & \qw
& \targ{} & \gate{R_Z(\phi_2)} & \targ{} & \cdots
& \targ{} & \gate{R_Z(\phi_d)} & \targ{} & \gate{H} & \qw \\
\lstick{$\ket{0}_{a_\mathcal{G}}^{\otimes m}$}
& \qw & \octrl{-1} & \qw & \octrl{-1} & \gate[wires=2]{U_\mathcal{G}}
& \octrl{-1} & \qw & \octrl{-1} & \gate[wires=2]{U_\mathcal{G}^\dagger}
& \octrl{-1} & \qw & \octrl{-1} & \cdots
& \octrl{-1} & \qw & \octrl{-1} & \qw & \qw \\
\lstick{$\ket{\psi}_{x}$}
& \qw & \qw & \qw & \qw & \qw
& \qw & \qw & \qw & \qw
& \qw & \qw & \qw & \cdots
& \qw & \qw & \qw & \qw & \qw
\end{quantikz}}}\right\}\,U_{V'}$
\end{adjustbox}
\caption{QSVT transformation of the block encoding $U_\mathcal{G}$ of $\mathcal{G}$ into a block encoding $U_{V'}$ of the degree-$(r-1)$ polynomial $V'(\mathcal{G})$.
QSVT uses $O(r-1)$ alternating queries to $U_\mathcal{G}$ and $U_\mathcal{G}^\dagger$.
The block-encoding signal ancilla $a_{\mathcal{G}}$ open-controls the CNOTs targeting the QSVT ancilla, which sandwich the phase rotations $R_Z(\phi_j)$. Post-selecting $a_\mathrm{qsvt} = a_\mathcal{G} = 0$ provides the correct polynomial transformation on $\mathcal{G}$.}
\label{fig:QSVT_Xgrid}
\end{appfigure}

\begin{appfigure}
\centering
\(\displaystyle
U_H=\begin{pmatrix} 0 & U_A^\dagger \\ U_A & 0 \end{pmatrix}
\)\par\smallskip
\begin{adjustbox}{max width=\columnwidth,center}
\begin{quantikz}[row sep=0.3cm, column sep=0.35cm]
\lstick{$\ket{0}_{a_H}$}
& \gate{X} & \octrl{1} & \ctrl{1} & \qw \\
\lstick{$\ket{\psi}_{U_A}$}
& \qw & \gate{U_A^\dagger} & \gate{U_A} & \qw
\end{quantikz}
\end{adjustbox}
\caption{Hermitian dilation of $U_A \longrightarrow U_H$.}
\label{fig:dilation}
\end{appfigure}

\begin{appfigure}
\centering
\begin{adjustbox}{max width=\columnwidth,center}
\begin{quantikz}[row sep=0.32cm, column sep=0.3cm]
\lstick{$\ket0_{a_\chi}$} & \qw & \octrl{2} & \qw & \cdots & \qw & \octrl{2} & \qw & \meter{0} \\
\lstick{$\ket0_{a_A}$} & \qw & \octrl{2} & \qw & \cdots & \qw & \octrl{2} & \qw & \meter{0} \\
\lstick{$\ket0_{a_\mathrm{qsvt}}$} & \qw & \octrl{2} & \qw & \cdots & \qw & \octrl{2} & \qw & \meter{0} \\
\lstick{$\ket0_{a_\mathcal{G}}^{\otimes m}$} & \qw & \octrl{1} & \qw & \cdots & \qw & \octrl{1} & \qw & \meter{0} \\
\lstick{$\ket0_{p}$} & \qw & \gate[wires=2]{\Phi_0} & \qw & \cdots & \qw & \gate[wires=2]{\Phi_d} & \qw & \meter{0} \\
\lstick{$\ket0_{\rm sgn}$} & \gate{H} & \qw & \qw & \cdots & \qw & \qw & \gate{H} & \meter{0}
\end{quantikz}
\end{adjustbox}\par\smallskip
\begin{adjustbox}{max width=\columnwidth,center}
\begin{quantikz}[row sep=0.12cm, column sep=0.3cm]
\lstick{$\scriptstyle a_\chi$} & \octrl{2} & \qw \\
\lstick{$\scriptstyle a_A$} & \octrl{2} & \qw \\
\lstick{$\scriptstyle a_\mathrm{qsvt}$} & \octrl{2} & \qw \\
\lstick{$\scriptstyle a_\mathcal{G}$} & \octrl{1} & \qw \\
\lstick{$\scriptstyle p$} & \gate[wires=2]{\Phi_k} & \qw \\
\lstick{$\scriptstyle \mathrm{sgn}$} & \qw & \qw
\end{quantikz}
\ = \
\begin{quantikz}[row sep=0.12cm, column sep=0.3cm]
\lstick{$\scriptstyle a_\chi$} & \octrl{2} & \qw & \octrl{2} & \qw \\
\lstick{$\scriptstyle a_A$} & \octrl{2} & \qw & \octrl{2} & \qw \\
\lstick{$\scriptstyle a_\mathrm{qsvt}$} & \octrl{2} & \qw & \octrl{2} & \qw \\
\lstick{$\scriptstyle a_\mathcal{G}$} & \octrl{1} & \qw & \octrl{1} & \qw \\
\lstick{$\scriptstyle p$} & \targ{} & \gate[wires=2]{\mathrm{R_{ZZ}}(2\phi_k)} & \targ{} & \qw \\
\lstick{$\scriptstyle \mathrm{sgn}$} & \qw & \qw & \qw & \qw
\end{quantikz}
\end{adjustbox}
\caption{Correct polynomial extraction of the QSVT block using the $\mathrm{sgn}$ ancilla.
\textit{Bottom:} the phase gadget $\Phi_k$ flips $p$ to $\ket1$ only on the success subspace $a_\chi=a_A=a_\mathrm{qsvt}=a_\mathcal{G}=0$, applies $\mathrm{R_{ZZ}}(2\phi_k)=e^{-i\phi_k Z\otimes Z}$ between $p$ and $\mathrm{sgn}$, and then uncomputes $p$. Thus $\mathrm{sgn}=0$ applies phase $+\phi_k$ and $\mathrm{sgn}=1$ applies $-\phi_k$.
\textit{Top:} reusing the same $\mathrm{sgn}$ qubit across all $\Phi_k$ coherently implements the QSVT sequences with phases ${+\phi_k}$ and ${-\phi_k}$, producing $U_{\Phi}=P(A)+iQ(A)$ and $U_{-\Phi}=P(A)-iQ(A)$. Preparing $\mathrm{sgn}$ in $\ket+$ and applying the final Hadamard gives the $\mathrm{sgn}=0$ branch
$\frac{U_{\Phi}+U_{-\Phi}}{2}=P(A)=\mathrm{Re}(U_{\Phi}).$
Post-selecting $\mathrm{sgn}=0$ together with $a_\chi=a_A=a_\mathrm{qsvt}=a_\mathcal{G}=p=0$ therefore removes the imaginary component $Q(A)$ and yields the desired real QSVT polynomial block $P(A)$.}
\label{fig:sgn-real-part}
\end{appfigure}

\begin{appfigure}
\centering
\(\displaystyle
\begin{aligned}
\mathrm{PREP}\ket{0}^{\otimes m}_{a_\mathcal{G}}
&= \frac{1}{\sqrt{\alpha_\mathcal{G}}}
\sum_{k=0}^{n-1}
\sqrt{2^{k} \frac{\Delta x}{2}}\ket{k}_{a_\mathcal{G}},\\
\phi_j&=2\arctan(r_j),\\
r_j&=2^{2^{m-1-j}/2}.
\end{aligned}
\)\par\smallskip
\begin{adjustbox}{max width=\columnwidth,center}
\begin{quantikz}[column sep=0.2cm]
\lstick{$\ket{0}_{a_\mathcal{G}}^{\otimes m}$} & \gate{\rm PREP} & \\
\end{quantikz}
\quad=\quad
\begin{quantikz}[row sep=0.1cm, column sep=0.2cm]
\lstick{$\ket{0}_{a_\mathcal{G}}$} & \gate{R_Y(\phi_0)} & \\
\setwiretype{n} \vdots & \vdots & \vdots\\
\lstick{$\ket{0}_{a_\mathcal{G}}$} & \gate{R_Y(\phi_{m-1})} &
\end{quantikz}
\end{adjustbox}
\caption{Preparation of the grid coefficient state using $R_Y$ rotations.
Shown for the special case $c_I=0$ and $n=2^m$ (e.g.\ the cell-centred grid $x_0=-L+\Delta x/2$), where the identity term of Fig.~\ref{fig:SEL_Xgrid} vanishes, $\alpha_\mathcal{G}=(N-1)\Delta x/2$, and the PREP register has $m=\log_2 n$ qubits; in general the identity term adds one branch and the register has $\lceil\log_2(n+1)\rceil$ qubits (Sec.~\ref{sec:resources}).}
\label{fig:PREP_Xgrid}
\end{appfigure}

\begin{appfigure}
\centering
\(\displaystyle
\mathcal{G}=\Big(x_0+\frac{(N-1)\Delta x}{2}\Big)I-\frac{\Delta x}{2}\sum_{k=0}^{n-1}2^kZ_k .
\)\par\smallskip
\begin{adjustbox}{max width=\columnwidth,center}
\begin{quantikz}[row sep=0.2cm,column sep=0.3cm]
\lstick{$\ket{0}^{\otimes m}_{a_\mathcal{G}}$}
& \gate{\mathrm{PREP}} & \muxctrl{1} & \cdots & \muxctrl{3} & \gate{\mathrm{PREP}^{\dagger}} & \qw \\
\lstick{$\ket{q_0}$}
& \qw & \gate{-Z_0} & \qw & \qw & \qw & \qw \\
& \vdots \setwiretype{n} & \setwiretype{n} & \push{\vdots} & \setwiretype{n} & \vdots \setwiretype{n} & \setwiretype{n} \\
\lstick{$\ket{q_{n-1}}$}
& \qw & \qw & \qw & \gate{-Z_{n-1}} & \qw & \qw
\end{quantikz}
\end{adjustbox}
\caption{LCU block encoding of the discretized position operator $\mathcal{G}$ using the $\mathrm{PREP}$-$\mathrm{SELECT}$-$\mathrm{PREP}^{\dagger}$ construction, with subnormalization $\alpha_\mathcal{G}=|c_I|+(N-1)\Delta x/2=\max_j|x_j|$, where $c_I=x_0+(N-1)\Delta x/2$. Branch $k<n$ of the PREP register applies $-Z_k$; the remaining branch applies $\mathrm{sgn}(c_I)\,I$, which acts only as a phase on the PREP register. In general the PREP register has $m=\lceil\log_2(n+1)\rceil$ qubits, one branch per Pauli term; Fig.~\ref{fig:PREP_Xgrid} shows PREP for the special case $c_I=0$.}
\label{fig:SEL_Xgrid}
\end{appfigure}

\end{document}